\documentclass{article}

\newcommand {\bl}{\begin{list}{}{\leftmargin 2em}}
\newcommand {\ite}{\item{}\hspace{-2em}}
\newcommand {\el}{\end{list}}
\newcommand {\beq}{\begin{equation}}
\newcommand {\eeq}{\end{equation}}

\newcommand{\bm}[1]{\mbox{\boldmath $#1$}}

\newcounter{saveeqn}
\newcommand{\alpheqn}{\setcounter{saveeqn}{\value{equation}}%
\stepcounter{saveeqn}\setcounter{equation}{0}%
\renewcommand{\theequation}{\mbox{\arabic{saveeqn}\alph{equation}}}}
\newcommand{\reseteqn}{\setcounter{equation}{\value{saveeqn}}%
\renewcommand{\theequation}{\arabic{equation}}}

\newcounter{EQ1}
\newcounter{EQ1A}
\newcounter{EQ2}
\newcounter{EQ3}
\newcounter{EQ4}
\newcounter{EQ5}
\newcounter{EQ6}

\newcounter{EQ7}
\newcounter{EQ8}
\newcounter{EQ8A}
\newcounter{EQ9}

\newcounter{EQ10}
\newcounter{EQ11}

\usepackage{graphicx, color}
\usepackage{exscale, latexsym, makeidx, newlfont}
\usepackage{amsfonts, amsbsy}
\usepackage[mathscr]{euscript}
\begin{document}

\begin{center}
\section*{Parameter estimation in differential equations: Mathematical foundations for satellite gravimetry, review and perspectives  }
\end{center}

\vspace{5mm}

\hspace*{-5mm}Peiliang Xu \\
 Disaster Prevention Research Institute, Kyoto University, Uji, Kyoto, Japan
 \\ pxu@rcep.dpri.kyoto-u.ac.jp

\vspace{5mm}

\begin{center}
\centerline{ \parbox{140mm}{ % \setlength{\baselineskip}{20pt}
{\bf Abstract}: {\sl Satellite gravimetry has become essential in many areas of earth science. However, the resolution of satellite gravitational models remains low at scales of a few hundred km and no gravity recovery methods can take full advantage of the unprecedented accuracy of satellite tracking measurements. We first provide a unified theoretical framework of parameter estimation in differential equations for satellite gravimetry and then briefly review the mathematical methods and modelings for computing the gravity field of the Earth from satellite tracking. More precisely, we focus on the collocation method, Kaula linear perturbations, two-point boundary value problems and orbit-energy-based methods. The orbit-energy methods are based on either Newtonian or  Hamiltonian mechanics. The numerical integration method is also included in this review, though it has been proved to be mathematically incorrect and physically not permitted. The reason is that it has become the standard method to routinely produce global gravitational models from satellite tracking data, which have gained the greatest possible impact and have been widely applied in many different areas of earth science. Because it is not clear how the incorrect mathematical foundation would affect gravity products from satellite tracking, we do not review any applications of these products. We then present a measurement-based perturbation theory to estimate the gravity field of the Earth, which can fully utilize both precise satellite orbits of arbitrary length and unprecedented  accuracy of satellite and inter-satellite tracking. The method is theoretically free of modeling errors, is capable of extracting any small forces from satellite and inter-satellite tracking data and provides a guarantee for high-precision high-resolution global gravity models. Finally, we assume a reference gravity model and derive local solutions to the Newton's nonlinear governing differential equations of satellite motion for scattered tracking data that can still be important in some applications.
 \vspace{2mm} \\ {\bf Key words:} parameter estimation in differential equations;
 Earth's gravity field; satellite gravimetry; measurement-based
perturbation; nonlinear differential equations; nonlinear Volterra's integral equations} } }

\end{center}

\vspace{4mm}

\section{Introduction}
Gravity and positioning constitute the foundations of geodesy, which is more generally defined as a science of accurate measurement of the shape and orientation, gravity field and spatiotemporal variations of the Earth. Interdisciplinary applications of geodesy are now also treated as part of geodesy. Gravity measurements are essential to determine the shape of the Earth, or more precisely, the flattening of the Earth (see e.g., Caputo 1967; Torge 2001) and the geoid (see e.g., Stokes 1849; Caputo 1967; Torge 2001), to develop the theory of isostatic compensation (Airy 1855; Pratt 1855; Dehlinger 1978; Torge 2001), to estimate free core nutation parameters (Florsch and Hinderer 2000) and for inertial navigation (see e.g., Grewal et al. 2001) and GNSS leveling or applied geodesy (see e.g., Kotsakis and Sideris 1999; Torge 2001). Gravity plays an important role in many other subject areas of earth science as well, for example, to understand the deformation of the lithosphere and even all mass deformation of the Earth under the concept of gravity tectonics (Ramberg 1967; de Jong and Scholten 1973; Dehlinger 1978), to confirm continental drift or plate motion with gravity as a plausible, even though not necessarily dominant, factor or evidence (de Jong and Scholten 1973; Dehlinger 1978), to determine mass distributions in geophysical prospecting (see e.g., Tsuboi and Fuchida 1937; Eckhardt 1940; Hammer 1945; Dehlinger 1978), and to invert for seafloor topography and study ocean circulation (see e.g., Delman and Landerer 2022; Jiang et al. 2024; Xu H. et al. 2024; Gou et al. 2025). On the other hand, gravity changes over time have been detected in association of local hydrological conditions (see e.g., Creutzfeldt et al. 2010), earthquakes and volcanic activities (see e.g., Barnes 1966; Chen et al. 1979; Yokoyama 1989; Tanaka et al. 2001; Okubo 2020) or more generally, mass redistributions (Torge 2001). For more applications, the reader may be referred to Crossley et al. (2013).

Gravity can be measured on land, at sea, in the air and/or from space. Although this paper is only focused on reviewing the mathematical foundations and modelings for the last topic, or more precisely, satellite-tracking gravimetry, it is worthwhile mentioning the pioneering works along the first three lines. The history of gravity measurement  started with Galileo’s laws of pendulum motion (in the early years of the seventeenth century but formally published much later in 1632 and 1638) (see e.g., Chapin 1998; Howarth 2007; Matthews 2014; Milsom 2018), which physically establishes the relationship of the period of oscillation, the length of the string and the gravity in one single formula. According to Howarth (2007) and Milsom (2018, p.32), the most accurate gravity measurement of 980.94 Gal was obtained by Huygens around 1660, though a rough estimate was believed to be derived earlier by Galileo himself. When Jean Richer was sent on an expedition to Cayenne of French Guiana, South America, in 1672–1673, he unexpectedly found that the pendulum clock is slower at Cayenne close to the equator than at Paris, indicating that the gravity is smaller at the equator and providing the observational evidence for the flattened or oblate Earth (see e.g, Lenzen and Multhauf 1966; Dehlinger 1978; Matthews 2014; Milsom 2018). This observational finding was well explained theoretically by Newton in his giant work {\em Principia} (see e.g., Poynting and Thomson 1907; Milsom 2018). More expeditions were sent to different parts of the world by the French Academy of Sciences to resolve the shape and size of the Earth. Among them, the expedition to Peru between 1735 and 1745, with Pierre Bouguer, Charles Marie de La Condamine and Louis Godin as the leaders, was the scientifically most successful. Bouguer measured gravity at three different altitudes with accurate pendulums, which led him to obtain the famous reduction of gravity now bearing in his name (see e.g., Poynting and Thomson 1907; Smallwood 2010; Milsom 2018).

When turning to gravity measurement at sea, we mention the pioneering work by Hecker, who worked under the directorship of Helmert at Potsdam, was mainly financially supported by the International Association of Geodesy and conducted the first gravity measurements in 1901, 1904 and 1909 (Hecker 1903; Bauer 1911; Dehlinger 1978; Milsom 2018). The most well-known sea gravity measurement expedition was Vening Meinesz's voyage to measure gravity with the modified  pendulums in a submarine of the Dutch Navy from 1923 to 1928, resulting in the finding of large negative gravity anomalies along the trench south of Java (Vening Meinesz 1929; LaCoste 1967; Dehlinger 1978; Tomoda 2010; Milsom 2018). For some more information on early marine gravity measurement, the reader may be referred to LaCoste (1967), Dehlinger (1978) and Tomoda (2010). With the invention of LaCoste and Romberg gravimeters, accurate measurement of gravity became possible. The first airborne gravity test in a U.S. Air Force KC-135 aircraft was reported by Thompson and LaCoste (1960) and soon followed by the second test in an aircraft with commercial equipment (Nettleton et al. 1960; LaCoste 1967). Instruments of gravity measurement have been revolutionarily advanced for about four centuries, ranging from pendulums up to 1970s (see e.g., Lenzen and Multhauf 1966; Dehlinger 1978), to (relative) gravimeters of spring type since the invention of LaCoste (1934, 1988) (see also Thompson and LaCoste 1960; Nettleton et al. 1960; LaCoste 1967), to absolute gravimeters from the early 1960s (Cook 1965, 1967a; Faller 1965), and finally to superconducting gravimeters starting about the same time as absolute gravimeters (see e.g., Crossley et al. 2013). Instruments for gravity gradient measurement have also evolved from torsion balances (see e.g., E\"{o}tv\"{o}s 1896; Poynting and Thomson 1907; Dehlinger 1978; Milsom 2018) to gradiometers comprising a set of accelerometers (see e.g., Jekeli 1993; Milsom 2018). Nevertheless, gravity measurement on a moving platform still remains challenged. This is particularly true, since precise acceleration measurement with accelerometers has never been properly addressed for more than one century. Only recently did Xu (2024) propose the concept of computerized accelerometers and re-define accelerometers as a combined system of physical sensing devices and reconstruction algorithms.

However, there is still about 12\% of the Earth’s land area with basically no terrestrial gravity data (either unavailable, too sparse or too inaccurate) (Pavlis et al. 2012). Even worse is that not all gravity data can be made available for free use. Pavlis et al. (2012) also reported that the area with 15 arc-minute area-mean gravity anomalies covered 42.9\% of the Earth’s total land area due to the agreement with the owners of  proprietary data. Obviously, terrestrial gravity plus sparse marine gravity data cannot lead to a precise gravity field of the Earth. Indeed, a high-precision high-resolution gravity field of the Earth demands a dense, global coverage of precise gravity data. It is only satellite gravimetry that can meet such a demand. One may argue that satellite altimetry could well provide a global coverage of oceans. Such an argument is valid only if prior knowledge about ocean dynamics is precisely available. Unfortunately, observables with satellite altimetry, the geoid and ocean dynamics are not separable, as can be seen, for example, in Rummel and Rapp (1977), Wunsch and Gaposchkin (1980) and Schrama (1989).

Satellite gravimetry has been topical for almost seven decades since the launch of the first artificial satellite Sputnik-1 by the former Soviet Union in 1957. With limited satellite tracking data before 2000, research interest in the topic was mainly within the geodetic community. The major achievements from scattered tracking data include the determination of the Earth's flattening (Buchar 1958; King-Hele et al. 1958), the pear component of the Earth's shape (O'Keefe et al. 1959) and first low degree and order satellite gravitational models (see e.g., Izsak 1963; Kaula 1963, 1966; Guier and Newton 1965). The demand for a global precise gravity field of the Earth called for dedicated satellite gravity missions, which were essential  to produce a global precise gravity field of the Earth, to meet the challenges of ocean dynamic topography, to provide a height datum and basis to replace time-consuming geodetic leveling together with global navigation satellite systems (GNSS), and to study physics of the interior of the Earth, mass re-distributions and fluid geospheres (see e.g., ESA 1996, 1999; NRC 1997; Dickey 2000). Thanks to the dedicated satellite gravity missions, namely, the Challenging Minisatellite Payload (CHAMP), the Gravity Recovery and Climate Experiment (GRACE) and the Gravity Recovery and Climate Experiment Follow-on (GRACE-FO) and the Gravity field and steady-state Ocean Circulation Explorer (GOCE), satellite gravimetry has become a key technique for almost all subject areas of earth sciences such as solid earth, hydrology, ocean, sea-level rise, glaciology, climate change and dynamic atmosphere (see e.g., NRC 1997; Tapley et al. 2019; Flechtner et al. 2021;  https://grace.jpl.nasa.gov/), even though its primary objective was to produce global gravitational models of the Earth. Actually, it is not exaggerated to say that geodesy nowadays is at the core of almost all earth-related disciplines. As of the time of writing (June 6, 2026), the key word ``satellite gravimetry'' produces about 7,130 publications according to google scholar (https://scholar.google.com/scholar?hl=en\&as\_sdt=0\%2C5\&q=\%22satellite+gravimetry\%22\&oq=). \\ This number of publications is not much different from 7,012 (as of October 2025) related to GRACE missions on the GRACE mission website (https://grace.jpl.nasa.gov/, access on June 6, 2026). Among these 7,130 publications, about 3,810 papers are related to terrestrial water storage, indicating that hydrological aspects have become the most important application area of satellite gravimetry. For the reason to be explained later in this paper, we will refrain from reviewing specific applications of satellite gravimetry here.

Satellite gravimetry is apparently very productive and extremely fruitful. However, it faces three critical and fundamental challenges: (i) all standard satellite gravitational models from satellite tracking are routinely computed by using the numerical integration method since 1970s (see e.g., Lerch et al. 1974; Long et al. 1989). Satellite gravitational models produced with measurements from CHAMP, GRACE and GRACE-FO missions have been widely applied in many areas of earth science, including but not limited to the subject areas mentioned above; actually, there are several thousands of publications of applications-nature in and far more beyond geodesy, as can be seen from the google scholar search in the above. However, the numerical integration method has been proved to be mathematically incorrect and physically not permitted (Xu 2009, 2018). He went further to propose mathematically rigorous methods to compute satellite gravity models from satellite tracking, though these methods have not yet been incorporated into operational practice. Most users of GRACE and GRACE-FO gravity products may not be aware of the fundamental incorrectness of the numerical integration method, nor of the extent to which this mathematical incorrectness may affect the conclusions of their research. This explains why we refrain from citing more {\em fruitful} applications from many different areas of earth science; (ii) the resolution of satellite gravity models from GRACE and GRACE-FO remains low at a scale of 300 to 500 km. From the point of view of mathematical modeling, there can be two main reasons for this: (a) such gravitational models have been computed with short arcs of orbit, limiting the increase of resolution and the capability of extracting small forces, because the effect of small forces cannot yet be sufficiently accumulated to be detectable; and (b) most methods depend on a reference model or small parameter perturbations to linearize a nonlinear dynamical system. As a result, such mathematical modelings are only valid within a certain span of time and will become divergent, which again will further limit their capability of extracting small forces from measurements. In addition to the most important issue of mathematical modeling, estimation methods such as regularization and non-conservative force modelings would add further aspects of challenge, in particular, with the increase of degrees and orders of the gravity field; and (iii) dedicated satellite gravity missions (CHAMP, GRACE, GRACE-FO and GOCE) are equipped with accelerometers to measure non-conservative forces. However, recent paradigm shift of accelerometers has demonstrated that raw data of acceleration from conventional accelerometers can be numerically incorrect and physically meaningless. Accelerometers have to be re-defined as a combined system of physical sensing devices and reconstruction algorithms (Xu 2024). It remains to be investigated to what extent this technological revolution will affect satellite gravity models.

This paper reviews the mathematical methods and modelings for reconstruction of the gravity field of the Earth from satellite tracking. We make no attempt to review any publications of applications-nature in many areas of earth science, because standard satellite gravity models such as those from GRACE and GRACE-FO missions are based on the fundamentally incorrect numerical integration  method. Readers interested in the applications and potential of satellite gravimetry may refer, for example, to NRC (1997), Dickey (2000), Tapley et al. (2019) and Flechtner et al. (2021). No attempt is made to review estimation methods such as regularization and non-conservative force modelings either. We will limit ourselves to the most important mathematical methods and modelings of satellite tracking in this paper, which fundamentally and critically decide the exactness and the extent of approximation of functional models of satellite tracking measurements. In the case of satellite gradiometry, we assume to directly measure the second gradients of the gravitational potential, though this assumption itself awaits computerized accelerometers or alike to be materialized. Observational equations of gravitational tensors are mathematically straightforward and GOCE mission will not be included in this review of mathematical modeling, though accelerometers are also one of concerns for not including GOCE mission and applications. The reader may be referred to Rummel (1986) and Rummel et al. (2011). 

The review is organized as follows. After the introduction with some brief historical information, Section 2 is to provide a unified framework of parameter estimation in differential equations for satellite gravimetry, and as a result, to reformulate the computation of satellite gravity models from satellite tracking under this framework. Section 3 is focused on the collocation or numerical differentiation methods. The numerical integration method will be included here as well, since it has been widely used by major institutions worldwide to produce satellite gravity products, and more importantly, since these products have found widest possible applications in many areas of earth science.  We will then review the conventional methods for satellite gravimetry, more precisely,  Kaula linear perturbations, two-point boundary value problems and orbit-energy-based methods in Section 4. Although short-arc and celestial mechanics methods have been developed and used after 2000, they will not be separately reviewed either, because their original theoretical or mathematical roots are either in two-point boundary value problems (see e.g., Ilk et al. 2005; Mayer-G\"{u}rr et al. 2005) or the numerical integration method (see e.g., Beutler et al. et al. 2010; J\"{a}ggi et al. 2016). To address the first two critical and fundamental challenges, finally in Section 5, we assume that low-earth-orbiting (LEO) gravity satellites can be precisely tracked continuously with GNSS, as is the case of the three dedicated satellite gravity missions, and present the measurement-based perturbation theory to solve the Newton's nonlinear differential equations of satellite motion, which is theoretically free of modeling errors and serves as a mathematical foundation for the determination of the high-precision high-resolution gravity field of the Earth from satellite and/or inter-satellite tracking. Since scattered tracking measurements can still be very important for some applications, and bearing in mind that almost all satellite tracking measurements were only scatteredly available in the early time of satellite gravimetry, we will construct local solutions to the nonlinear differential equations of satellite motion based on a reference gravity model as well. So far as a sufficient number of satellite tracking data are collected and linked to any of these solutions, one can then statistically estimate the gravity model of the Earth.

\section{Parameter estimation in differential equations as a unified mathematical foundation for satellite gravimetry from satellite tracking}
Consider the following one-dimensional (1D) differential equation:
\beq \label{basicDiffEQ1D} \dot{x} = f( x(t), t, \mathbf{p}), \eeq where $\dot{x}$ stands for the derivative of the function $x(t)$ with respect to the time $t$ and $\mathbf{p}$ is a vector of unknown parameters in the differential equation (\ref{basicDiffEQ1D}). Given a set of $n$ measurements collected at time epochs $(t_1, t_2, ..., t_n)$, namely, $\mathbf{y} = \{ y_1(x(t_1)), y_2(x(t_2)), ..., y_n(x(t_n))\}$, we want to estimate the unknown parameters $\mathbf{p}$ from the measurements $\mathbf{y}$. If different tools or instruments are involved at a certain epoch $t_i$, we have more than one measurement, denoted as $\mathbf{y}_i(x(t_i))$,  instead of just one single $y_i(x(t_i))$. In practice, one may assume that the system state $x(t)$ is directly measured and obtain $\mathbf{y} = \{ x(t_1), x(t_2), ..., x(t_n)\}$. Without loss of generality and without technical difficulty, the 1D differential equation (\ref{basicDiffEQ1D}) can be readily extended to a vector of differential equations of $\mathbf{x}(t)$ with the unknown parameters $\mathbf{p}$. Estimation of unknown parameters in differential equations has been often encountered in many areas of science and engineering, for example, in mathematics (see e.g., Ritt 1919; Gronwall 1919; Coddington and Levinson 1955; Howland and Vaillancourt 1961; Riley et al. 1967; Van Domselaar and Hemker 1975; Varah 1982; Hu et al. 2015), in statistics (see e.g., Ramsay et al. 2007; Wang and Enright 2013), in chemical engineering (see e.g., Peterson 1962; Van den Bosch and Hellinckx 1974; Hwang et al. 1978; Law and Sharma 1997; Linga et al. 2006; Sarode et al. 2015), in physics (see e.g., Dickinson and Gelinas 1976; Scitovski and Juki\'{c} 1996) and in ecology, biology and genome (see e.g., Benson 1979; Moles et al. 2003; Peifer and Timmer 2007; Brewer et al. 2008), just to name some of application areas.

Given a set of satellite tracking data $\mathbf{y}$, satellite gravimetry is essentially equivalent to solving for the unknown parameters in the following Newton's nonlinear differential equations:
\beq \label{NewtonLAW} \ddot{\mathbf{x}}(t) = \mathbf{a}_E(t, \mathbf{x}(t),
\dot{\mathbf{x}}(t), \mathbf{p}) + \mathbf{a}_M(t, \mathbf{x}(t), \dot{\mathbf{x}}(t),
\mathbf{p}_M), \eeq (see e.g., Xu 2008, 2018), where $\mathbf{x}(t)$ is the position vector of the satellite in the inertial reference frame, $\ddot{\mathbf{x}}(t)$ stands for the second derivatives of $\mathbf{x}(t)$ with respect to time $t$,  $\mathbf{a}_E(t, \mathbf{x}(t), \dot{\mathbf{x}}(t), \mathbf{p})$ is the
Earth's gravitational attraction exerted on the satellite with the unknown parameters
$\mathbf{p}$ to be estimated, and $\mathbf{a}_M(t, \mathbf{x}(t), \dot{\mathbf{x}}(t), \mathbf{p}_M)$ represents all other forces such as solid earth and ocean tides, atmospherical drag, solar radiation pressure and third-body effects (see e.g., Taff 1985; Reigber 1989; Jekeli 1999; Seeber 2003), $\mathbf{p}_M$ is the vector of the unknown parameters of these force models. Although both terms $\mathbf{a}_E(t, \mathbf{x}(t),
\dot{\mathbf{x}}(t)(t), \mathbf{p})$ and  $\mathbf{a}_M(t, \mathbf{x}(t), \dot{\mathbf{x}}(t),
\mathbf{p}_M)$ are clearly of different physical sources, they are of no essential difference from the point of view of statistical estimation. Thus, in what follows, without loss of generality, we will limit ourselves to the Earth's gravitational attraction. The Newton's nonlinear differential equations (\ref{NewtonLAW}) can be accordingly simplified as: \alpheqn \beq \label{NewtonGrav} \ddot{\mathbf{x}}(t) = \mathbf{a}_E(t, \mathbf{x}(t), \mathbf{p}), \eeq where $\mathbf{a}_E(t, \mathbf{x}(t), \mathbf{p})$ is the gradient of the geopotential and is given by
 \beq \label{accGrav} \mathbf{a}_E(t, \mathbf{x}(t), \mathbf{p})
= - \frac{GM}{r^3}\mathbf{x}(t) + \frac{\partial T(r,\mathbf{p})}{\partial \mathbf{x}(t)}. \eeq
Here $GM$ is the product of the Earth's mass $M$ and the gravitational
constant $G$ and is often denoted by $\mu=GM$, $r=\|\mathbf{x}(t) \|$, and $T(r,\mathbf{p})$ is the disturbing potential of the Earth's gravity field, which is given in the non-inertial
earth-fixed (ECEF) reference frame as follows: \beq \label{TPotential} T(r,\mathbf{p}) =
\frac{GM}{r}\sum \limits^{\infty} \limits_{l=2} \sum \limits_{m=0} \limits^l
\left( \frac{R}{r}\right)^l[ C_{lm} \cos(m\lambda) + S_{lm} \sin(m\lambda) ]
P_{lm}(\cos\theta), \eeq \reseteqn\setcounter{EQ1}{\value{equation}}(see e.g., Groves 1960; Kaula 1966; Heiskanen and Moritz 1967), where $R$ is the mean radius of the Earth, $C_{lm}$ and $S_{lm}$ are the
normalized, dimensionless harmonic coefficients, $\lambda$ and $\theta$ are the
longitude and colatitude of the satellite, respectively, and $P_{lm}(t)$ is the
normalized associated Legendre function. The parameter vector $\mathbf{p}$ consists of all the unknown harmonic coefficients $C_{lm}$ and $S_{lm}$. Since the number of satellite tracking data is always finite, one has to practically always replace the infinity in (\ref{TPotential}) with a maximum degree and
order $N_{\scriptsize \textrm{max}}$. Note, however, that the coordinate transformation of the position
vector is required, because (\ref{NewtonGrav}) and (\ref{TPotential}) are
formulated in different reference systems (see e.g., Taff 1985; Jekeli 1999). Without confusion, we will assume that the reader is well aware of the transformation and will not emphasize this point in the remainder of this review. If we denote $\mathbf{v}(t)=\dot{\mathbf{x}}(t)$ and $\mathbf{z}(t)=[\mathbf{x}^T(t), \mathbf{v}^T(t)]^T$, then the second order differential equations (\ref{NewtonGrav}) can be equivalently rewritten as the first order differential equations: \beq \label{2ndto1stDiffE} \dot{\mathbf{z}}(t) = \left[ \begin{array}{c} \mathbf{v}(t) \\ \mathbf{a}_E(t, \mathbf{x}(t), \mathbf{p}) \end{array} \right]. \eeq The right hand side of (\ref{2ndto1stDiffE}) will be denoted, without confusion, as the vector functions $\mathbf{f}(t, \mathbf{z}(t), \mathbf{p})$ for convenience and conciseness of notations.

In the early time of satellite gravimetry (see e.g., Brouwer and Clemence 1961;
Kaula 1966; Hagihara 1972; Taff 1985), one often worked with the six Keplerian orbital elements, namely, $\mathbf{K}=[a, e, \omega, i, \Omega, \overline{M}]$, with each orbital element of $\mathbf{K}$ standing for the semi-major axis of the orbital ellipse, the eccentricity, the argument of the perigee, the inclination of the orbital plane, the longitude of the ascending node and the mean anomaly, respectively. In this case, the second order differential equations (\ref{NewtonGrav}) of motion of artificial satellites can alternatively be rewritten mathematically as the first order differential equations in orbital elements as follows:
\alpheqn \beq \label{EqKeplerA} \frac{da}{dt} =
\frac{2}{na}\frac{\partial T(\mathbf{K},\mathbf{p})}{\partial \overline{M}}\eeq \beq \label{EqKeplerE} \frac{de}{dt} =
\frac{1-e^2}{na^2e}\frac{\partial T(\mathbf{K},\mathbf{p})}{\partial \overline{M}} -
\frac{(1-e^2)^{1/2}}{na^2e}\frac{\partial T(\mathbf{K},\mathbf{p})}{\partial \omega} \eeq \beq
\label{EqKepleromega} \frac{d\omega}{dt} = -
\frac{\cos\,i}{na^2(1-e^2)^{1/2}\sin\,i}\frac{\partial T(\mathbf{K},\mathbf{p})}{\partial i} +
\frac{(1-e^2)^{1/2}}{na^2e} \frac{\partial T(\mathbf{K},\mathbf{p})}{\partial e} \eeq \beq \label{EqKepleri}
\frac{di}{dt} = \frac{\cos\,i}{na^2(1-e^2)^{1/2}\sin\,i}\frac{\partial T(\mathbf{K},\mathbf{p})}{\partial
\omega} - \frac{1}{na^2(1-e^2)^{1/2}\sin\,i} \frac{\partial T(\mathbf{K},\mathbf{p})}{\partial \Omega} \eeq \beq
\label{EqKeplerOmega} \frac{d\Omega}{dt} =
\frac{1}{na^2(1-e^2)^{1/2}\sin\,i}\frac{\partial T(\mathbf{K},\mathbf{p})}{\partial i} \eeq \beq
\label{EqKeplerM} \frac{d\overline{M}}{dt} = n - \frac{1-e^2}{na^2e} \frac{\partial T(\mathbf{K},\mathbf{p})}{\partial e} -
\frac{2}{na}\frac{\partial T(\mathbf{K},\mathbf{p})}{\partial a}
\eeq\reseteqn\setcounter{EQ2}{\value{equation}}after the sign convention of Kaula (1966), where $n$ is the mean motion, and $T(\mathbf{K},\mathbf{p})$ is the disturbing potential of the Earth but in the Keplerian orbit elements. For other (equivalent) versions of  Lagrange's planetary equations (\arabic{EQ2}), the reader is referred to, for example, Brouwer and Clemence (1961), Kaula (1966), Taff (1985) and Seeber (2003).

Now let us further assume that the satellite tracking data $\mathbf{y}$ is of a weighting matrix $\mathbf{W}$. If the weighted least squares (LS) method is applied to estimate the gravity field of the Earth, then satellite gravimetry from satellite tracking can be finally formulated mathematically as the following optimization model: \beq \label{satGravOptim} \textrm{min:} \hspace{2mm} \{ \mathbf{y} - \mathbf{y}[\mathbf{x}(t)] \}^T\mathbf{W}\{ \mathbf{y} - \mathbf{y}[\mathbf{x}(t)] \}, \eeq subject to the constraints (\ref{NewtonGrav}), or equivalently, (\ref{2ndto1stDiffE}) or (\arabic{EQ2}), where $\mathbf{y}$ is a vector of satellite tracking observables, $\mathbf{y}[\mathbf{x}(t)]$ is the computed vector of satellite tracking data at the orbit $\mathbf{x}(t)$ of the satellite. Note that $\mathbf{x}(t)$ is an implicit function vector of the unknown parameters $\mathbf{p}$. The estimation problem (\ref{satGravOptim}) of satellite gravimetry is essentially an optimization model with the equality constraints in terms of nonlinear differential equations (\ref{NewtonGrav}) or (\arabic{EQ2}), which looks formally different from conventional geodetic estimation in the sense that the equality constraints are now given in the form of nonlinear differential equations but not our familiar algebraic equations. However, it is this fundamental difference that makes satellite gravimetry very difficult and there is no easy solution to it. If a high-resolution gravity field is desirable, instead of using the LS method of (\ref{satGravOptim}), one would have to apply regularization to estimate $\mathbf{p}$. Because the differential equations (\ref{NewtonGrav}) or equivalently (\arabic{EQ2}) are highly nonlinear, geodesists have almost always used linear perturbations and the numerical integration method to determine the gravity field of the Earth from satellite tracking (see e.g., Anderle 1965; Kaula 1966; Lerch et al. 1974; Long et al. 1989). More about these methods will be discussed later.

In principle, one may first assume an unknown initial value $\mathbf{x}(t_0)$, solve the differential equations  (\ref{NewtonGrav}), obtain a formal solution $\mathbf{x}(t, \mathbf{p}, \mathbf{x}(t_0))$, compute the values $\mathbf{y}[\mathbf{x}(t, \mathbf{p}, \mathbf{x}(t_0))]$, and finally estimate the unknown force parameters $\mathbf{p}$ from the satellite tracking data $\mathbf{y}$ by minimizing the objective function of (\ref{satGravOptim}) or by applying regularization methods. However, since the differential equations  (\ref{NewtonGrav}) are nonlinear, it is generally not possible to obtain the analytical solution $\mathbf{x}(t, \mathbf{p}, \mathbf{x}(t_0))$. Therefore, any mathematically rigorous solution to satellite gravimetry from satellite tracking can only be numerically iterative up to the present. Actually, this is also exactly the case in many other areas of science and engineering involved with parameter estimation in differential equations. Because the objective function is nonlinear, one may have to resort to evolutionary algorithms of stochastic nature (see e.g., Moles et al. 2003), implying that one can, at most, obtain an improved solution to (\ref{satGravOptim}). If computation cost is not an issue of concern, one may apply interval-based deterministic global optimization algorithms (see e.g., Hansen 1992; Xu 2003) to find the exact global optimal solution to (\ref{satGravOptim}).

If a number of satellites are involved, we have the same number of constraints of type (\ref{NewtonGrav}), or equivalently  (\ref{2ndto1stDiffE}) or (\arabic{EQ2}). By collecting all the tracking observables to and/or between the satellites into the vector, again denoted by $\mathbf{y}$ with a weighting matrix $\mathbf{W}$ without confusion, in this case, the optimization model (\ref{satGravOptim}) becomes \beq \label{multiSatGravOptim} \textrm{min:} \hspace{2mm} \{ \mathbf{y} - \mathbf{y}[\mathbf{x}_1(t),\mathbf{x}_2(t),...,\mathbf{x}_k(t)] \}^T\mathbf{W}\{ \mathbf{y} - \mathbf{y}[\mathbf{x}_1(t),\mathbf{x}_2(t),...,\mathbf{x}_k(t)] \}, \eeq with each $\mathbf{x}_i(t)\, (i=1,2,...,k)$ representing the trajectory of the {\em i}th satellite at the observed discrete time epochs and satisfying its corresponding differential equation of type (\ref{NewtonGrav}), or equivalently (\ref{2ndto1stDiffE}) or (\arabic{EQ2}). Each element of $\mathbf{y}[\mathbf{x}_1(t),\mathbf{x}_2(t),...,\mathbf{x}_k(t)]$ may be involved with one or more artificial satellites. For example, inter-satellite tracking observables have to be expressed as the differences of the two individual solutions. If the system of observational equations becomes unstable, in particular, in the case of gravity models with a high degree and order, regularization or similar methods should be used instead to accurately and stably estimate the unknown harmonic coefficients $\mathbf{p}$.

\section{The collocation and numerical integration methods}
We will now briefly review and comment on two most popular methods to estimate the unknown parameters in the differential equations (\ref{basicDiffEQ1D}) or equivalently (\ref{2ndto1stDiffE}) from measurements, namely the collocation method and the numerical integration method. Both of them have been widely applied in many areas of science and engineering. In particular, bearing in mind that the latter has been routinely used to produce standard gravitational models from satellite tracking, we will have to spend more effort to explain and comment on this method, mathematically and physically.

\subsection{The collocation or numerical differentiation method}
The collocation method is based on numerical differentiation and is one of the most commonly used methods to estimate the unknown parameters in differential equations. The basic idea is to use basis functions to fit the data, compute the derivatives of the fitted curves, turn the differential equations into the observation equations and finally apply the LS method to estimate the unknown parameters (see e.g., Van den Bosch and Hellinckx 1974; Varah 1982; Tjoa and Biegler 1991; Scitovski and Juki\'{c} 1996; Reubelt et al. 2003; Ramsay et al. 2007; Brewer et al. 2008; Sarode et al. 2015). With the differential equation (\ref{basicDiffEQ1D}) in mind, this class of methods always assumes direct measurements on $x(t)$, namely, $\mathbf{y} = \{ x(t_1), x(t_2), ..., x(t_n)\}$. One can then use some orthogonal basis functions or cubic/kernel functions to fit these data points. Let us assume that the fitted function is denoted by $\overline{x}(t)$ and is given by
 \beq \label{basisFit} \overline{x}(t) = \sum \limits_{i=1} \limits^n b_i g_i(t), \eeq where $b_i$ are the given coefficients obtained by fitting the data $\mathbf{y}$ and $g_i(t)$ are the basis functions.

Applying the derivative operator to $\overline{x}(t)$ of (\ref{basisFit}), we can then rewrite the differential equation (\ref{basicDiffEQ1D}) as follows:
\beq \label{derivedObservables} \dot{\overline{x}}(t_i) = \sum \limits_{i=1} \limits^n b_i \dot{g}_i(t_i) \approx f( x(t_i), t_i, \mathbf{p}), \eeq or equivalently in the form of observational equations, \beq \label{derivedObservablesFinal} \dot{\overline{x}}(t_i)  + \epsilon_i = f( x(t_i), t_i, \mathbf{p}), \eeq for $i=\{1,2,...,n\}$. If the computed derivatives are assumed to be of a weighting matrix $\mathbf{W}$, we can estimate the unknown parameters $\mathbf{p}$ by solving the following unconstrained nonlinear (weighted) LS problem:
\beq \label{collocationOptim} \textrm{min:} \hspace{2mm} \{ \overline{\mathbf{y}} - \mathbf{f}( x(t_i), t_i, \mathbf{p}) \}^T\mathbf{W}\{ \overline{\mathbf{y}} - \mathbf{f}( x(t_i), t_i, \mathbf{p}) \}, \eeq where both $\overline{\mathbf{y}}$ and $\mathbf{f}( x(t_i), t_i, \mathbf{p})$ are given by:
  $$ \overline{\mathbf{y}} = [ \dot{\overline{x}}(t_1), \dot{\overline{x}}(t_2), ..., \dot{\overline{x}}(t_n) ]^T, $$ and
  $$ \mathbf{f}( x(t_i), t_i, \mathbf{p}) = [ f( x(t_1), t_1, \mathbf{p}), f( x(t_2), t_2, \mathbf{p}), ..., f( x(t_n), t_n, \mathbf{p}) ]^T. $$

{\em Remark 1:} The advantage of the collocation method is easy to implement without the need to solve the most difficult part of the problem, namely, finding the solution to the nonlinear differential equation (\ref{basicDiffEQ1D}), if the system state $x(t)$ is directly measured. Nevertheless, if the measurements only contain partial information on $x(t)$, it is not always possible/easy to recover $x(t_i)$ from such partial information. In this case, one will encounter difficulty in applying the collocation method to such partial information.

\begin{figure}
  \centering
  \includegraphics[width=110mm,height=20mm]{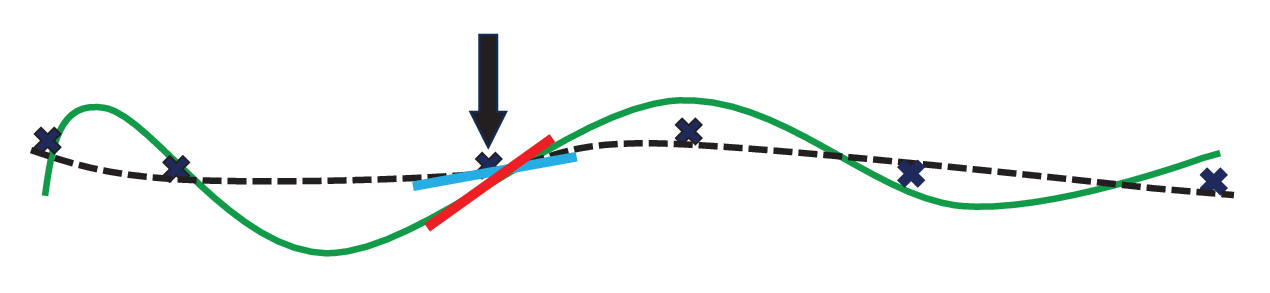}
  \caption{An illustrative data set with marks $\times$. The fitted and true curves of the data are represented  in the dashed black and solid green lines, respectively. Also plotted in this illustrative figure are the derivatives at one of the marked points (indicated by the black arrow), where the derivative of the fitted curve is shown in light blue and that of the true curve is shown in red, respectively.}
  \label{computDerivatives}
\end{figure}

{\em Remark 2:} The main problem with the collocation method is twofold: (i) if data is sparse and scattered, the fitted curve may be quite different from its true curve. To illustrate this issue and its impact on the computed derivatives, we show a set of scattered data in Fig.~\ref{computDerivatives}, together with the fitted curve in dashed black line and the true curve in green line, respectively. It can be clearly seen from this illustrative figure that the derivatives $\dot{\overline{x}}(t_i)$ computed with the fitted curve can be significantly different from their corresponding true values, as illustrated at one of the data points (compare the light blue and red lines in Fig.~\ref{computDerivatives}); (ii) computing the derivatives from noisy data is well known to be ill-posed (see e.g., Xu et al. 2021; Xu 2024). In other words, if data is very dense, unless regularization is applied, the computed first and/or second derivatives from the noisy data can be simply incorrect numerically, implying that any further inference about or estimation of the parameters with them can become meaningless; and (iii) even in this case, a naive application of the weighted LS method to (\ref{derivedObservablesFinal}) can lead significant bias in the estimated parameters of $\mathbf{p}$ (Xu et al. 2014; Xu 2019), depending on the noise level of measurements.

\subsection{The numerical integration method}\label{dynamInt}
The numerical integration method of satellite gravimetry was first hinted at in the 1965 U.S. Naval Weapons Laboratory technical report by Anderle (1965) and then developed by Riley et al. (1967). After the method was formally documented in the NASA Goddard Space Flight Center Technical Reports by Lerch et al.
(1974) and Long et al. (1989) (see also Ballani 1988; Reigber 1989; Montenbruck and Gill 2000;
Beutler et al. 2010), it has become the standard method to compute the Earth's gravity models from satellite tracking, in particular, the gravitational models from GRACE and GRACE-FO missions, which have been widely applied in many areas of earth science. Actually, the method had been independently developed and well known before Anderle (1965) as the sensitivity analysis method in many other areas of science and engineering (see e.g., Dickinson and Gelinas 1976; Hwang et al. 1978; Law and Sharma 1997; Linga et al. 2006; Peifer and Timmer 2007; Wang and Enright 2013). If we further traced back to history, it is interesting to see that the first two papers on this topic were published in {\em Ann Math} by Ritt (1919) and Gronwall (1919) more than one century ago, which was followed in the mathematics book of Coddington and Levinson (1955).

The numerical integration method for satellite gravimetry, as formulated in (\ref{2ndto1stDiffE}),
starts with the following differential equations:
\beq \label{PartialSmatr} \dot{\mathbf{S}}(t, \mathbf{z}(t), \mathbf{p}) = \frac{\partial \mathbf{f}(t, \mathbf{z}(t), \mathbf{p})}{\partial \mathbf{z}^T} \mathbf{S}(t, \mathbf{z}(t), \mathbf{p}) + \frac{\partial \mathbf{f}(t, \mathbf{z}(t), \mathbf{p})}{\partial \mathbf{p}^T}, \eeq where $\mathbf{S}(t, \mathbf{z}(t), \mathbf{p})$ is the matrix of the partial derivatives of $\mathbf{z}(t)$ with respect to $\mathbf{p}$, namely,
$$  \mathbf{S}(t, \mathbf{z}(t), \mathbf{p}) = \frac{\partial \mathbf{z}(t)}{\partial \mathbf{p}^T}. $$
The differential equations (\ref{PartialSmatr}), as first derived by Ritt (1919) and followed by Gronwall (1919),
 are {\em apparently extremely} useful, since they {\em seemly provide a unique way} to compute the partial derivatives of the satellite orbit with respect to the force parameters $\mathbf{p}$. Indeed, they have become the standard method in satellite tracking gravimetry since the publication of Anderle (1965) (see also  Lerch et al. 1974; Long et al. 1989) and have been routinely applied to produce gravitational models from GRACE and GRACE-FO for widest possible applications in many areas of earth science.  We should like to note that the differential equations of this type are of the secondary or derived nature of the original differential equations (\ref{2ndto1stDiffE}). From this point of view, they do not add any new information nor new insight beyond the original differential equations (\ref{2ndto1stDiffE}) (Xu 2018). In other words, if one would not be able to use the original differential equations (\ref{2ndto1stDiffE}), together with a given but unknown initial value $\mathbf{z}(t_0)$, to solve the difficult, highly nonlinear satellite gravimetry problem of estimating the force parameters $\mathbf{p}$ from satellite tracking data $\mathbf{y}$, it would be equally helpless to illusively solve the same mathematical problem with the aid of the secondary or derived equations (\ref{PartialSmatr}).

Unfortunately, the history of the numerical integration or sensitivity method really follows the unrealistic illusion. More precisely, to solve the differential equations for $\mathbf{S}(t, \mathbf{z}(t), \mathbf{p})$, one requires {\em the} initial value of $\mathbf{S}(t, \mathbf{z}(t), \mathbf{p})$ at the initial time epoch $t_0$, which is always assumed to be zero, namely,
\beq \label{SmatrT0} \mathbf{S}(t_0, \mathbf{z}(t_0), \mathbf{p}) = \left. \mathbf{S}(t, \mathbf{z}(t), \mathbf{p})\right|_{t=t_0} = \mathbf{0}, \eeq as can be first seen in the mathematical literature (Gronwall 1919; Coddington and Levinson 1955), then more or less independently in the literature of satellite gravimetry (see e.g, Riley et al. 1967; Ballani 1988; Montenbruck and Gill 2000) and in other areas of science and engineering (see e.g., Dickinson and Gelinas 1976; Hwang et al. 1978; Law and Sharma 1997; Linga et al. 2006; Peifer and Timmer 2007; Wang and Enright 2013). For some more historical details about the initial value (\ref{SmatrT0}), the reader is referred to Xu (2009, 2018).

For the reader to better understand the fundamental erroneousness of the numerical integration method, we note that the solution $\mathbf{z}(t)$ to the differential equations (\ref{2ndto1stDiffE}) must be the functions of $\mathbf{p}$, no matter whether we explicitly write it in terms of $\mathbf{p}$, namely, $\mathbf{z}(t, \mathbf{p})$. The same logic applies to $\mathbf{z}(t_0)$ or equivalently $\mathbf{z}(t_0, \mathbf{p})$. More precisely, both $\mathbf{z}(t)$ and $\mathbf{z}(t_0)$ contains the physical information in terms of $\mathbf{p}$. 
In the remainder of this section, we will review the proofs that $\mathbf{z}(t_0)$ are indeed the functions of $\mathbf{p}$ and further that the initial values of $\mathbf{S}(t, \mathbf{z}(t), \mathbf{p})$ cannot be equal to zero. New proof results that are not available in Xu (2009, 2018) will also be provided. In what follows, for conciseness, we will use the conventional notations of $\mathbf{z}(t)$ and $\mathbf{z}(t_0)$.

\begin{figure}[th]
  \centering
  \includegraphics[width=110mm,height=90mm]{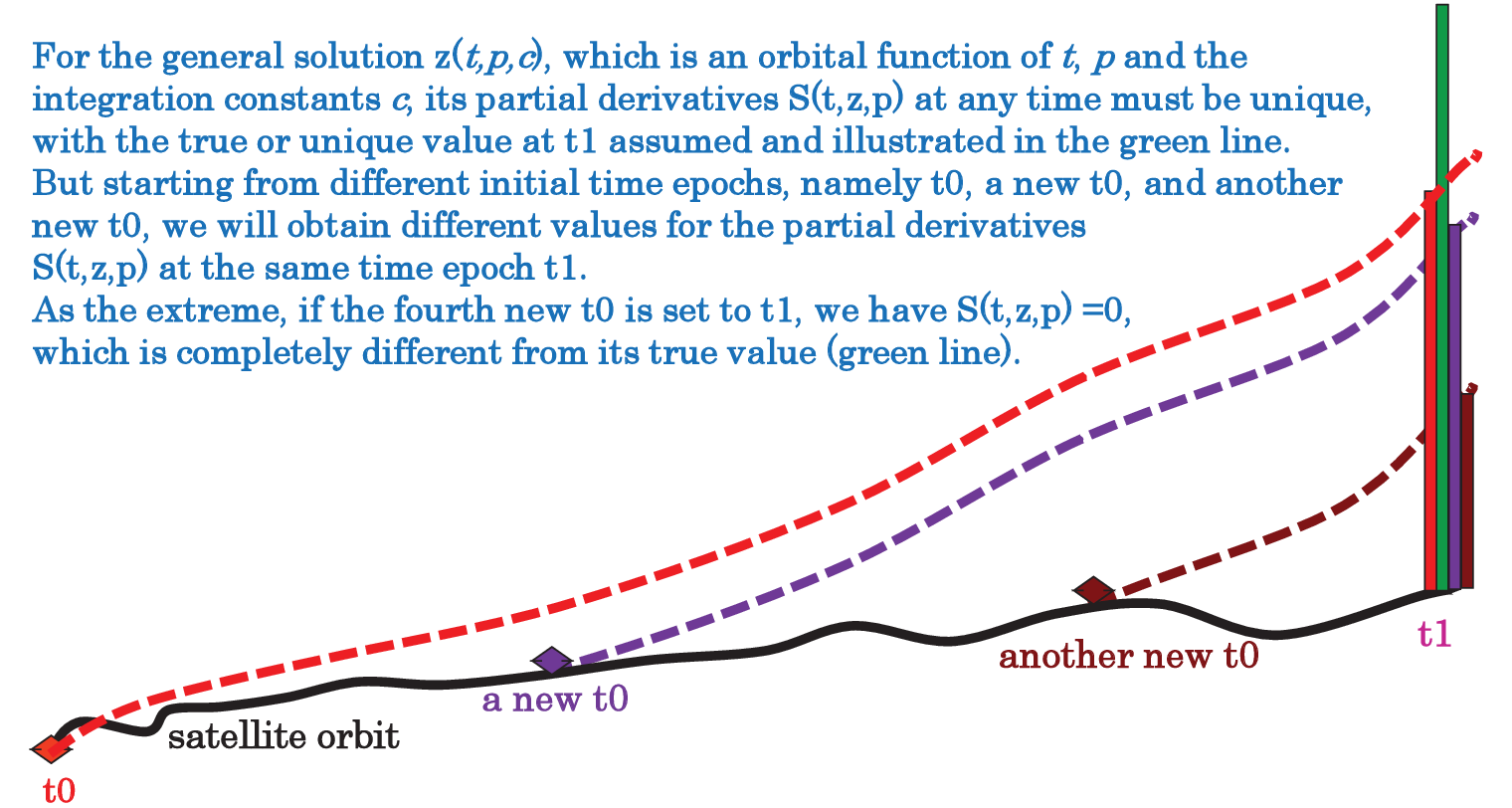}
  \caption{An illustrative example of $\mathbf{S}(t, \mathbf{z}(t), \mathbf{p})$ for a differential equation with unknown parameters $\mathbf{p}$, which is unique for a particular satellite orbit (shown in the black line). The true value of $\mathbf{S}(t, \mathbf{z}(t), \mathbf{p})$ at $t_1$ for the differential equation is assumed and shown in the green line. Starting from each of the initial epochs $t_0$, $\textrm{a new}\,\, t_0$ and $\textrm{another new}\,\, t_0$, we will obtain different values of $\mathbf{S}(t, \mathbf{z}(t), \mathbf{p})$ at $t_1$ under the zero initial condition (\ref{SmatrT0}), which are shown in red, purple and brown lines at $t_1$, respectively. As an extreme, if the fourth initial epoch is set to $t_1$, we have $\mathbf{S}(t_1, \mathbf{z}(t_1), \mathbf{p})=0$. These results contradict the fact that $\mathbf{S}(t, \mathbf{z}(t), \mathbf{p})$ is unique for the particular trajectory of satellite. Note that this figure only serves to illustrate the different values of $\mathbf{S}(t, \mathbf{z}(t), \mathbf{p})$ with different initial time epochs but does not imply in any sense that the functions $\mathbf{S}(t, \mathbf{z}(t), \mathbf{p})$ monotonically increase. }
  \label{diffSmatrT1}
\end{figure}

Although the differential equations (\ref{PartialSmatr}) of the partial derivatives
$\mathbf{S}(t, \mathbf{z}(t), \mathbf{p})$ are mathematically correct (Ritt 1919), Xu (2009, 2018) proved that the initial values of $\mathbf{S}(t, \mathbf{z}(t), \mathbf{p})$ cannot be equal to zero in satellite gravimetry from satellite tracking along different lines: (i) the counter-examples are constructed to prove that $\mathbf{S}(t_0, \mathbf{z}(t_0), \mathbf{p})$ cannot be equal to zero; (ii) given an initial value $\mathbf{z}(t_0)$, one can formally obtain the corresponding satellite orbit or solution $\mathbf{z}(t, \mathbf{z}(t_0), \mathbf{p})$. This initial value $\mathbf{z}(t_0)$ is nothing but a point of the solution $\mathbf{z}(t, \mathbf{z}(t_0), \mathbf{p})$. Because of the arbitrariness of $t_0$, if (\ref{SmatrT0}) would be true, then $\mathbf{S}(t, \mathbf{z}(t), \mathbf{p}) = \mathbf{0}$ must have been true along the whole orbit as well. This is obviously scientifically impossible, since it implies that the satellite orbit would not be a function of $\mathbf{p}$, further contradicting our starting differential equations (\ref{2ndto1stDiffE}) with $\mathbf{p}$ and even worse, implying that satellite orbits would be physically irrelevant to the gravitational forces; (iii) if (\ref{SmatrT0}) would be true, and bearing in mind the arbitrariness of initial time $t_0$ (say $t_{01}, t_{02}, ..., t_{0m}$), then one could use as arbitrarily many $t_0$ as possible by setting $m$ to any large number to form a set of equations: \beq \label{SmatrParam} \left. \begin{array}{c} \mathbf{S}(t_{01}, \mathbf{z}(t_{01}), \mathbf{p}) = \mathbf{0} \\
 \mathbf{S}(t_{02}, \mathbf{z}(t_{02}), \mathbf{p}) = \mathbf{0} \\ \vdots \\ \mathbf{S}(t_{0m}, \mathbf{z}(t_{0m}), \mathbf{p}) = \mathbf{0} \end{array} \right\} \eeq The equation system (\ref{SmatrParam}) would imply that one could immediately solve for the gravity parameters $\mathbf{p}$ by only setting $m$ to a sufficiently large number without the need to observe or track satellites. This is again scientifically impossible and contradictory to the fact that satellite observables are indeed required to determine the gravity field of the Earth; and finally, (iv) for a particular satellite orbit or equivalently, for the solution of a particular satellite motion, the partial derivatives $\mathbf{S}(t, \mathbf{z}(t), \mathbf{p})$ must be unique at any given time epoch, say $t_1$. However, starting with a different initial time $t_{0i}$, one would then solve the differential equations (\ref{PartialSmatr}) under the zero initial condition (\ref{SmatrT0}) at the initial time $t_{0i}$ and obtain a different value $\mathbf{S}(t_1, \mathbf{z}(t_1), \mathbf{p})$ at $t_1$, which clearly contradicts the uniqueness of $\mathbf{S}(t, \mathbf{z}(t), \mathbf{p})$ at the time $t_1$, as illustrated in Fig.~\ref{diffSmatrT1}. To summarize, we have mathematically proved that $\mathbf{S}(t_0, \mathbf{z}(t_0), \mathbf{p})$ cannot be equal to zero but are the functions of $\mathbf{p}$. So far as the initial condition for the original differential equations $(\ref{2ndto1stDiffE})$ is given, the initial condition for the secondary or derived differential equations (\ref{PartialSmatr}) is automatically fixed and cannot be arbitrarily given, as in the case of (\ref{SmatrT0}). This also explains why we use the language of {\em the} initial condition but not an initial condition for (\ref{PartialSmatr}).

Because the source of the zero initial condition is in the {\em Ann Math} paper by Gronwall (1919) (see also Coddington and Levinson 1955), we will now directly disprove this zero initial condition mathematically and make some comments on this paper.

Gronwall (1919) assumed a set of $n$ differential equations of type (\ref{2ndto1stDiffE}) but with only a single parameter, namely, $\mathbf{p}=a$. Under the initial condition $\mathbf{z}(t_0)$, Gronwall formally represented the solutions to the differential equations as $\mathbf{z}(t, \mathbf{z}(t_0), a)$. Up to this moment, in principle, Gronwall should be able to go ahead to compute the derivative $\mathbf{S}(t_0, \mathbf{z}(t_0), a)$ directly from $\mathbf{z}(t, \mathbf{z}(t_0), a)$ and further represent it as follows: \beq \label{GronwallSmatr0} \mathbf{S}(t_0, \mathbf{z}(t_0), a) = \left. \frac{d\mathbf{z}(t, \mathbf{z}(t_0), a)}{da}\right|_{t_0} = \frac{\partial \mathbf{z}(t_0, \mathbf{z}(t_0), a)}{\partial \mathbf{z}^T(t_0)} \frac{\partial \mathbf{z}(t_0)}{\partial a} + \frac{\partial \mathbf{z}(t_0, \mathbf{z}(t_0), a)}{\partial a}. \eeq However, instead of computing $\mathbf{S}(t_0, \mathbf{z}(t_0), a)$ in (\ref{GronwallSmatr0}), Gronwall chose to solve the differential equations of type (\ref{PartialSmatr}) with an initial condition. As a result, Gronwall  simply set the initial value $\mathbf{S}(t_0, \mathbf{z}(t_0), a)$ to zero, namely,
\beq \label{GronwallS0} \mathbf{S}(t_0, \mathbf{z}(t_0), a) = \left. \frac{d\mathbf{z}(t, \mathbf{z}(t_0), a)}{da}\right|_{t_0} = \mathbf{0} \eeq  without any proof or arguments.  The condition (\ref{GronwallS0}) logically speaks that the original differential equations do not contain the parameter $a$ because of the arbitrariness of $t_0$, which clearly contradicts the starting assumption. Furthermore, since $\mathbf{S}(t_0, \mathbf{z}(t_0), a)$ has been uniquely determined by the original differential equations under the initial condition $\mathbf{z}(t_0)$, it cannot be arbitrarily given, as is done in (\ref{GronwallS0}), unfortunately. In other words, the initial condition for the derived differential equations of type (\ref{PartialSmatr}) must be derived on the basis of the solution to the original differential equations. The arbitrarily given condition (\ref{GronwallS0}) contradicts the natural results of derivatives in (\ref{GronwallSmatr0}). Any information that cannot be derived from the original differential equations is not permitted to be imposed on the derived differential equations (\ref{GronwallSmatr0}). Even worse, if one would be allowed to have the zero initial  condition (\ref{GronwallS0}), this would be equivalent to:
\beq \label{GronwallEQ} \frac{\partial \mathbf{z}(t_0, \mathbf{z}(t_0), a)}{\partial \mathbf{z}^T(t_0)} \frac{\partial \mathbf{z}(t_0)}{\partial a} + \frac{\partial \mathbf{z}(t_0, \mathbf{z}(t_0), a)}{\partial a} = \mathbf{0}. \eeq

In general, the sensitivity vector $\mathbf{S}(t, \mathbf{z}(t), a)$ may or may not be functions of $a$. We will now further demonstrate the contradictions of the initial condition (\ref{GronwallEQ}) to the original differential equation and thus prove again the mathematical incorrectness of the initial condition (\ref{GronwallEQ}) from three different logical reasonings, though one of them has been presented in Xu (2009, 2018). First of all, if $\mathbf{S}(t, \mathbf{z}(t), a)$ is not a vector function of $a$, then it can be at most a vector function of $t$. Because of the arbitrariness of $t_0$, the initial condition (\ref{GronwallEQ}) in this case is mathematically equivalent to $\mathbf{S}(t, \mathbf{z}(t), a)=\mathbf{0}$ for an arbitrary time $t$, further implying mathematically that $\mathbf{S}(t, \mathbf{z}(t), a)$ is completely not sensitive to $a$ and that $\mathbf{z}(t)$ does not contain any information on $a$ physically. This is a clear contradiction to the original differential equation with the unknown parameter $a$; secondly, if $\mathbf{S}(t, \mathbf{z}(t), a)$ is a vector function of $a$, the initial condition (\ref{GronwallEQ}) constitutes a system of equations. Because of arbitrariness of the original differential equations, it is possible that the solution to each equation of this system does not exist. However, the original differential equation has no relevance to this possibility, indicating that the initial condition (\ref{GronwallEQ}) can contradict the original differential equations in this case; and finally, even if the existence of solution to each of the equations (\ref{GronwallEQ}) can be assumed and if such a solution can be further assumed to be unique, because we have a set of $n$ equations in (\ref{GronwallEQ}), it is possible to obtain $n$ values for $a$. The original differential equation does not permit this scenario at all. Practically, this further implies that we could even solve for different or contradictory values of $a$ without any measured data. In particular, such a contradiction has no relevance to any randomness in the data (if any) at the logical level. To conclude, the initial zero condition (\ref{GronwallS0}) or (\ref{GronwallEQ}) is nothing more than a contradictory claim and cannot be permitted mathematically.

Actually, the reader can readily follow the five lines in this section, as applied to $\mathbf{S}(t_0, \mathbf{z}(t_0), \mathbf{p})$, and disprove the zero initial condition of Gronwall (1919). We may like to note that some mathematicians might be aware of this contradictory claim, as may be inferred from Van Domselaar and Hemker (1975). In fact, Van Domselaar and Hemker (1975) directly assumed that $\mathbf{S}(t_0, \mathbf{z}(t_0), \mathbf{p})$ are the functions of $\mathbf{p}$ and are known in advance. Nevertheless, this can only be possible in simulations but not possible without first knowing the original functions $\mathbf{z}(t,\mathbf{p})$.

We may further like to note that Milbert and Jekeli (2023) attempted to comment on the zero initialization of the sensitivity matrix. Although Xu (2009, 2018) proved that the zero initial condition (\ref{SmatrT0}) cannot be true, Milbert and Jekeli (2023) did not show anything wrong mathematically in all the proof procedures of Xu (2009, 2018). Instead, they chose to assume that the initial condition was independent of $\mathbf{p}$ or equivalently that the zero initial condition (\ref{SmatrT0}) was true (compare their assumption, {\em i.e.},  their equation (5)) and went on to prove that the same zero initial condition (\ref{SmatrT0}) (compare their result, {\em i.e.}, their equation (11)) was correct. This is logically incorrect, because it is equivalent to proving that A is correct by assuming that A is true. Their reasoning is clearly circular and does not constitute a proof. The remainder of Milbert and Jekeli (2023) just repeated the same logic mistake; because of this logic mistake, their comments on the zero initialization of the sensitivity matrix are logically invalid. Therefore, the work of Milbert and Jekeli (2023) is irrelevant to whether (\ref{SmatrT0}) is correct or not and requires no further comment here.

To conclude the discussion of the numerical integration method, I would like to quote a general statement by Taff (1985, p.322) as follows: \vspace{2mm} \\ \hspace*{4mm} \begin{minipage}{13cm}
{\em Not having a method that allows the solution of complex problems (or even simple problems) is no excuse for using unjustified techniques.} \end{minipage} \vspace{2mm} \\even though Taff said it under a different context. Probably as for the numerical integration method, ``{\em unjustified techniques}'' may be more appropriately replaced with ``{\em mathematically incorrect methods}''.

Finally, before closing this section, we show an example for which we can obtain an analytical solution to differential equations with unknown parameters. Let us assume the linear differential equations with unknown parameters $\mathbf{p}$ as follows:
\beq \label{ODELinear} \dot{\mathbf{x}}(t) = \mathbf{F}(t, \mathbf{p}) \mathbf{x}(t) \eeq under the initial condition $\mathbf{x}(t_0)$. Based on linear Kalman filtering theory, we can readily write the analytical solution of the linear differential equations (\ref{ODELinear}) with the aid of the concept of state transition matrices: \beq \label{ODESolut} \mathbf{x}(t) = \mathbf{\Phi}(t,t_0,\mathbf{p}) \mathbf{x}(t_0), \eeq where $\mathbf{\Phi}(t,t_0,\mathbf{p})$ is the state transition matrix. By substituting $\mathbf{x}(t)$ of (\ref{ODESolut}) into (\ref{ODELinear}), we have $$\dot{\mathbf{\Phi}}(t,t_0,\mathbf{p}) \mathbf{x}(t_0) =\mathbf{F}(t, \mathbf{p}) \mathbf{\Phi}(t,t_0,\mathbf{p}) \mathbf{x}(t_0), $$ which is equivalent to \alpheqn  \beq \label{transPHI}
\dot{\mathbf{\Phi}}(t,t_0,\mathbf{p}) = \mathbf{F}(t, \mathbf{p}) \mathbf{\Phi}(t,t_0,\mathbf{p}), \eeq because of the arbitrariness of $\mathbf{x}(t_0)$. On the other hand, by setting $t$ to $t_0$ in (\ref{ODESolut}) and considering the arbitrariness of $\mathbf{x}(t_0)$, we obtain the initial condition \beq \label{initialConditionPHI} \mathbf{\Phi}(t_0,t_0,\mathbf{p}) = \mathbf{I} \eeq\reseteqn\setcounter{EQ1A}{\value{equation}}to solve for the transition matrix of (\ref{transPHI}), where $\mathbf{I}$ is an identity matrix. Note, however, that the initial value $\mathbf{x}(t_0)$ for (\ref{ODELinear}) is implicitly a function of the unknown parameters $\mathbf{p}$, while the initial condition (\ref{initialConditionPHI}) for $\mathbf{\Phi}(t,t_0,\mathbf{p})$ of (\ref{transPHI}) is completely free of $\mathbf{p}$.

\section{Kaula linear perturbations, two-point boundary value problems and the orbit energy-based methods}
The first artificial satellite Sputnik-1 stands as a fundamental technological revolution in the history of geodesy and starts the era of precise measurement of the Earth from space, both geometrically and physically. Satellite gravimetry has since become one of the scientifically most fruitful topics in geodesy and found widest possible applications in many areas of earth science, some of which may even hardly be imagined then (see e.g., NRC 1997; Dickey 2000). A number of methods were soon proposed after the launch of Sputnik-1 to meet the immediate demand of processing satellite tracking data and to determine the gravity field of the Earth from space measurements. In this section, we will briefly review three major classes of methods developed in the early time of satellite gravimetry, more precisely, Kaula linear perturbations, two-point boundary value problems and the orbit-energy-based methods.

\subsection{Kaula linear perturbations}
Perturbation methods were the most fruitful and played a dominant role in the determination of the Earth's gravity field from a very limited number of satellite tracking data  in the 1950s through the late 1970s (see e.g., Buchar 1958; Brouwer 1959; King-Hele and Merson 1959; O'Keefe et al. 1959; Groves 1960; Brouwer and Clemence 1961; Kozai 1959, 1962; Cook 1961, 1963; Izsak 1963; Guier and Newton 1965; Kaula 1963, 1966; Hagihara 1972; Gaposchkin 1974; Lambeck and Coleman 1983; Taff 1985; Seeber 2003; Kloko\v{c}n\'{i}k et al. 2013). Although perturbations can be applied to rectangular coordinates (see e.g., Brumberg 1978), almost all perturbation methods with important applications directly work with orbital elements, as can be seen from the above references, with Kaula linear perturbations as its most outstanding example, in particular, in the geodetic community. Thus, in what follows, we will briefly review Kaula linear perturbation theory. More technical details can be found, for example, in Kaula (1966) and Seeber (2003).

Since the gravitational field of the Earth consists of two parts, namely, the central force term and the small forces in terms of the disturbing potential $T(r,\mathbf{p})$ of (\ref{TPotential}), the basic idea of Kaula linear perturbations is to follow the small parameter perturbation theory and work out the linear perturbations of the orbital elements $\mathbf{K}$ with respect to $T(r,\mathbf{p})$, or more precisely, the (small) unknown harmonic coefficients $(C_{lm},\,S_{lm})$. As the first step, one has to transform the disturbing potential $T(\mathbf{r},\mathbf{p})$  of (\ref{TPotential}) from spherical coordinates into the six orbital elements $\mathbf{K}$ as follows: \alpheqn
\beq \label{TPotentialKOrbits} T(\mathbf{K},\mathbf{p})  =
\sum \limits^{N_{max}} \limits_{l=2} \sum \limits_{m=0} \limits^l T_{lm}, \eeq where \begin{eqnarray} \label{potentialTlm}
T_{lm} & = & \frac{\mu a_e^l}{a^{l+1}} \sum \limits^{l} \limits_{p=0} F_{lmp}(i) \sum \limits^{\infty} \limits_{q=-\infty} G_{ipq}(e) S_{impq}(\omega, \overline{M}, \Omega, \theta) \nonumber \\
  & = & \sum \limits^{l} \limits_{p=0} \sum \limits^{\infty} \limits_{q=-\infty} T_{lmpq}, \end{eqnarray}
\beq \label{SFun} S_{impq}(\omega, \overline{M}, \Omega, \theta)   =  \left[ \begin{array}{c} C_{lm} \\ -S_{lm} \end{array} \right]_{l-m\,\, \textrm{odd}}^{l-m\,\, \textrm{even}} \cos(\psi_{lmpq}) + \left[ \begin{array}{c} S_{lm} \\ C_{lm}  \end{array} \right]_{l-m\,\, \textrm{odd}}^{l-m\,\, \textrm{even}} \sin(\psi_{lmpq}), \eeq
 \beq \label{funPHI} \psi_{lmpq} = (l-2p)\omega+(l-2p+q)\overline{M} + m(\Omega-\theta), \eeq\reseteqn\setcounter{EQ3}{\value{equation}}$a_e$ is the equatorial radius of the Earth, $\theta$ is Greenwich Sidereal Time, $F_{lmp}(i)$ and $G_{ipq}(e)$ are the inclination and eccentricity functions, respectively, which are given, for example, in Kaula (1966) and Seeber (2003).

After all these preparations, the second step is to insert $T_{lmpq}$ of (\ref{potentialTlm}) into Lagrange's planetary equations (\arabic{EQ2}) and then integrate both sides of the equations with $T_{lmpq}$ from $t_0$ to $t$. At this key step of integration, one has to treat all the orbital elements $[a, e, \omega, i, \Omega, n]$ as if they were constants, though they are really the functions of time. Also at this step of integration, one further treats $\dot{\omega}$, $\dot{\overline{M}}$ (or equivalently $n$), $\dot{\Omega}$ and $\dot{\theta}$ as if they were constants as well when integrating $S_{impq}(\omega, \overline{M}, \Omega, \theta)$. Finally, one can obtain the linear perturbation for each element of $\mathbf{K}$ due to $T_{lmpq}$ as the function of the small (unknown) harmonic coefficients $(C_{lm},\,S_{lm})$, namely, $\Delta \mathbf{K}_{lmpq}=[\Delta a_{lmpq}, \Delta e_{lmpq}, \Delta \omega_{lmpq}, \Delta i_{lmpq}, \Delta \Omega_{lmpq}, \Delta \overline{M}_{lmpq}]$, where each of $\Delta \mathbf{K}_{lmpq}$ is inversely proportional to the terms below:
\beq \label{resonancePHI} \dot{\psi}_{lmpq} = (l-2p)\dot{\omega}+(l-2p+q)\dot{\overline{M}} + m(\dot{\Omega}-\dot{\theta}). \eeq We should like to note that in the derivation of the perturbations due to $T_{lmpq}$, the (extra) constant terms at the time $t_0$ can be simply neglected from $\Delta \mathbf{K}_{lmpq}$, because they can be readily merged into the initial values of the orbital elements. By summing up all the linear perturbation terms $\Delta \mathbf{K}_{lmpq}$, one can then obtain the final linear perturbations for all the orbital elements due to the disturbing potential $T(\mathbf{K},\mathbf{p})$ of (\ref{TPotentialKOrbits}). Since $\Delta \mathbf{K}_{lmpq}$ has been well documented in, for example, Kaula (1966, formula 3.76) and Seeber (2003, formula 3.119), we will not repeat them here.

{\em Remark 3:} Perturbations played a key role both in the determination of the gravity field of the Earth in the early time of satellite gravimetry and in deriving and understanding physical behaviours of satellite orbits, even though some physical behaviours are just artifacts of perturbation mathematics. It is clear from $S_{impq}(\omega, \overline{M}, \Omega, \theta)$ of (\ref{SFun}) that if $\psi_{lmpq}$ of (\ref{funPHI}) is equal to zero, then the effect of the corresponding disturbing potential $T_{lm}$ of (\ref{potentialTlm}) would behave linearly with time, which is called secular perturbations. In this case, $m$ must be equal to $0$, implying that secular perturbations are attributed to only the zonal coefficients (Seeber 2003). If $\psi_{lmpq} = (l-2p)\omega$ in (\ref{funPHI}), the perturbation of the corresponding disturbing potential $T_{lm}$ would be of long-period nature. Since the indices $p$ and $q$ of (\ref{funPHI}) change quickly and since the term $m(\Omega-\theta)$ represents the change of $m$ cycles per day, these two terms contribute to short-periodical perturbations. Actually, the secular perturbations or the averaging effect of the orbital elements were often used to determine the gravity field of the Earth (see e.g., Kaula 1963, 1966; Cook 1967b). In particular, if $\dot{\psi}_{lmpq}$ of (\ref{resonancePHI}) is equal to zero, $\Delta \mathbf{K}_{lmpq}$ would become infinity, which is called {\em resonance}. This phenomenon is physically not real but a consequence of coordinate definition (Kaula 1966; Xu 2008). We may also like to note that critical inclination is physically not real either (see e.g., Taff 1985; Xu 2008), though it was a hot topic of many publications (see e.g., Kaula 1966; Coffey et al. 1986; Jupp 1988; Breiter and Elipe 2006).

{\em Remark 4:} Strictly speaking, linear perturbations should be a direct result of rigorously solving the nonlinear Lagrange's planetary equations (\arabic{EQ2}) up to linear approximation. However, as can be seen in the above, Kaula linear perturbation theory has to make two assumptions for integration: (i) the orbital elements $[a, e, \omega, i, \Omega, n]$ are constant; and (ii) $\dot{\omega}$, $\dot{\overline{M}}$ (or equivalently $n$), $\dot{\Omega}$ and $\dot{\theta}$ are constant as well. It is not clear how these extra assumptions would affect the perturbations of the orbital elements, since a mathematically rigorous linear perturbation should only be based on the assumption of the small harmonic coefficients. Nevertheless, since the numerical integration method was published, Kaula linear perturbation theory has basically not been used to produce standard gravity models of the Earth from satellite tracking, though we now know that the numerical integration method itself is a mathematically incorrect foundation, as can be seen in Section~\ref{dynamInt} (see also Xu 2009, 2018). We may further note that linear perturbations of this kind mathematically belong to small parameter perturbations and will get divergent with the increase of time.

\subsection{Two-point boundary value problems}
Two-point boundary value problems (BVP) were first formulated by Schneider (1968) for orbit determination in geodesy, with a brief outline for use as a new method to determine the gravity field of the Earth from satellite tracking. The two-point BVP theory has since been further developed by Schneider (1984, 2006) himself and applied for global gravity field recovery, in particular, by the research group of Professor H. Ilk (see e.g., Ilk et al. 2005, 2008; Mayer-G\"{u}rr et al. 2005).

More precisely and without loss of generality for applications to satellite gravimetry, we may rewrite the differential equations (\ref{NewtonGrav}) as follows: \alpheqn \beq \label{twoPointBVP} \ddot{\mathbf{x}}(t) = \mathbf{f}(t, \mathbf{x}(t)), \eeq subject to the boundary conditions at the two boundary points, namely, \beq \label{twoPointEnd} t_0,\,\,\mathbf{x}(t_0); \hspace{2mm}\textrm{and}\hspace{2mm} t_1, \,\, \mathbf{x}(t_1). \eeq\reseteqn\setcounter{EQ4}{\value{equation}}The solution to the two-point BVP is well known to be: \alpheqn
\beq \label{twoPointSolutST} \mathbf{x}(t) = \mathbf{x}(t_0) + \frac{\mathbf{x}(t_1)-\mathbf{x}(t_0)}{T} (t-t_0) -  \int_{t_0}^{t_1} K(t,s)\mathbf{f}(s, \mathbf{x}(s))ds, \eeq (see e.g., Schneider 1968; Stakgold 2000), where $T=t_1-t_0$, and the kernel function $K(t,s)$ is given as follows:
\beq \label{kernelFunST} K(t, s) = \left\{ \begin{array}{c}
(t-t_0)(t_1-s)/T, \hspace{2mm} \textrm{if} \hspace{1.5mm} t \leq s \\
(s-t_0)(t_1-t)/T, \hspace{2mm} \textrm{if} \hspace{1.5mm} s \leq t \end{array} \right. \eeq\reseteqn\setcounter{EQ5}{\value{equation}}Given the force functions $\mathbf{f}(t, \mathbf{x}(t))$ and the boundary conditions (\ref{twoPointEnd}), one can then use the solution (\arabic{EQ5}) to compute the orbit of a satellite.

Instead of using the solution (\arabic{EQ5}) for orbit determination, Schneider (1968, 1984, 2006) (see also Ilk et al. 2005, 2008) re-scaled the time interval $[t_0,\,t_1]$ to $[0,\,1]$ and rewrote the solution (\ref{twoPointSolutST})  as
\alpheqn \beq \label{twoPointSolut} \mathbf{z}(\eta) = \mathbf{x}(t_0) + (\mathbf{x}(t_1)-\mathbf{x}(t_0)) \eta - T^2 \int_{0}^1 K(\eta, \xi)\mathbf{f}(t_0+T\xi, \mathbf{z}(\eta))d\xi, \eeq where $\eta=(t-t_0)/T$, $\xi=(s-t_0)/T$,
 \beq \label{zFunx} \mathbf{z}(\eta) = \mathbf{x}(t_0+T\eta), \eeq
and the kernel or Green function $K(t, s)$ of (\ref{kernelFunST}) becomes \beq \label{kernelFun} K_0(\eta, \xi) = \left\{ \begin{array}{c}
\eta (1- \xi), \hspace{2mm} \textrm{if} \hspace{1.5mm} \eta \leq \xi \\
\xi(1-\eta), \hspace{2mm} \textrm{if} \hspace{1.5mm}\xi \leq \eta \end{array} \right. \eeq\reseteqn\setcounter{EQ6}{\value{equation}}For gravity field recovery, Schneider (1968) further rewrote the kernel function $K_0(\eta, \xi)$ as a bilinear representation of the eigenfunctions:
\alpheqn \beq \label{kernelBilinear} K_0(\eta, \xi) = 2 \sum \limits_{n=1} \limits^{\infty} \frac{\sin(n\pi \eta)\sin(n\pi \xi)}{(n\pi)^2}, \eeq (see also Stakgold 2000), and obtained the solution in terms of eigenfunctions: \beq \label{twoPointSolutEigen} \mathbf{z}(\eta) = \mathbf{x}(t_0) + (\mathbf{x}(t_1)-\mathbf{x}(t_0)) \eta + \sum \limits_{n=1} \limits^{\infty} c_n\sin(n\pi \eta), \eeq where the coefficients $c_n$ are determined by the force functions and given below:
\beq \label{coeffSolut} c_n = - \frac{2T^2}{(n\pi)^2} \int_{0}^1 \sin(n\pi \xi)\mathbf{f}(t_0+T\xi, \mathbf{z}(\xi))d\xi. \eeq\reseteqn\setcounter{EQ7}{\value{equation}}

From the mathematical point of view, the solutions (\ref{twoPointSolutST}), (\ref{twoPointSolut}) and (\ref{twoPointSolutEigen}) to the differential equations (\ref{twoPointBVP}) under the boundary conditions (\ref{twoPointEnd}) are equivalent for orbit determination. Schneider (1968, 1984) showed that the coefficients $c_n$ can be estimated from satellite tracking data. If there exist some unknown parameters $\mathbf{p}$ in the force functions, the right hand side of (\ref{twoPointBVP}) will become $\mathbf{f}(t, \mathbf{x}(t), \mathbf{p})$. As a result, the estimated coefficients $c_n$ can be linked to the unknown force parameters $\mathbf{p}$ through (\ref{coeffSolut}) by substituting $\mathbf{f}(t, \mathbf{x}(t))$ with $\mathbf{f}(t, \mathbf{x}(t), \mathbf{p})$. In other words, Schneider (1968, 1984, 2006) demonstrated that the orbit determination through the formulation of the two-point BVP can be adapted for gravity field recovery from satellite tracking data of short arcs (see also Ilk et al. 2005, 2008; Mayer-G\"{u}rr et al. 2005). We note, however, that the use of short arcs is not capable of extracting small forces, implying that a high-precision high-resolution gravitational model is not possible with such short-arc data.  Actually, one can also directly apply the solution (\ref{twoPointSolutST}) or (\ref{twoPointSolut}) to the determination of the gravity field of the Earth without the use of (\ref{coeffSolut}). Instead, satellite tracking data can always be more conveniently linked to the solution. Because the differential equations (\ref{twoPointBVP}) are highly nonlinear and include  the unknown parameters $\mathbf{p}$, one would have to estimate $\mathbf{p}$ iteratively from measurements. A slight inconvenience of using the two-point BVP theory for gravity field recovery may be that the kernel function $K(t, s)$ of (\ref{kernelFunST}) will have to change with $t_1$ and $T$.

\subsection{The orbit-energy-based methods}
The orbit-energy-based methods for measuring the gravity field of the Earth were also proposed soon after the launch of the first artificial satellite. They were developed almost at the same time as the two-point BVP theory by Schneider (1968), though these two types of methods were roughly proposed about one decade later than the perturbation methods. The orbit-energy-based methods were first advanced by Bjerhammar (1967, 1969) and likely, independently, by Wolff (1969), since Wolff (1969) did not refer to the work of Bjerhammar (1967). The idea was quickly taken up by Hotine and Morrison (1969), with reference to the work of Bjerhammar. While Bjerhammar (1967, 1969) put emphasis on working out the measurement of the geopotential with one satellite disturbed by the Earth and other celestial bodies such as the Sun and the Moon, Wolff (1969) came up with the idea of measuring the geopotential with a pair of satellites. The latter method can be more powerful, since one can collect precise relative measurements of range rates between two satellites.  To prepare for the realization of twin satellites, one tracking the other, and in particular, thanks to the launch of GRACE and GRACE-FO missions, the method of Wolff (1969) has attracted much more interest, modification, extension and real applications (see e.g., Morrison 1970; Comfort 1974; Jekeli 1999; Visser et al. 2003; Gerlach et al. 2003; Han et al. 2006) than the two-point BVP theory, though their popularity is completely not comparable with the (incorrect) numerical integration method.

If a satellite (of unit mass) is ideally treated as a freely falling test mass in space and if conservative potential energies other than that of the Earth are negligible, its total mechanical energy (i.e. energy per unit mass) consists of two parts: kinetic energy and potential energy, namely, \beq \label{satEnergy} E = \|\dot{\mathbf{x}}_i(t)\|^2/2 - V(\mathbf{x}_i(t)), \eeq  where $E$ is the total mechanical energy, $\|\dot{\mathbf{x}}_i(t)\|$ is the velocity magnitude of the satellite, the subscript $i$ stands for the Earth-centered inertial reference frame (ECI), and $V(\mathbf{x}_i(t))$ is the gravitational potential of the Earth represented in the ECI. There is a minus sign before the geodetic potential energy because of the difference in sign convention between physics and geodesy (Bjerhammar 1967, 1969; Jekeli 1999). The satellite mechanical energy $E$ of (\ref{satEnergy}) is conserved and is equal to a constant, as can be seen from its derivative with respect to time:
\begin{eqnarray} \label{EConservation}  \frac{dE}{dt} & = & \dot{\mathbf{x}}_i^T(t)\ddot{\mathbf{x}}_i(t) - [\nabla V(\mathbf{x}_i(t))]^T \dot{\mathbf{x}}_i(t) \nonumber \\
 & = & \dot{\mathbf{x}}_i^T(t) \nabla V(\mathbf{x}_i(t)) - [\nabla V(\mathbf{x}_i(t))]^T \dot{\mathbf{x}}_i(t) = 0, \end{eqnarray} with $\ddot{\mathbf{x}}_i(t)=\nabla V(\mathbf{x}_i(t))$ in mind.

We may note that: (i) the energy  conservation law (\ref{satEnergy}) physically speaks that the larger the geopotential at the position of a satellite, the faster the satellite flies there, or alternatively, the smaller the geopotential at the position of a satellite, the slower the satellite flies; and (ii) $V(\mathbf{x}_i(t))$ of (\ref{satEnergy}) is a scalar and physically invariant, no matter what reference system is used to compute the geopotential. We represent $V(\mathbf{x}_i(t))$ of (\ref{satEnergy}) in the ECI, which is only the function of position $\mathbf{x}_i(t)$ but not of the explicit time $t$. However, if we represent the invariant geopotential in the ECEF, then $V(\mathbf{x}_i(t))$ of (\ref{satEnergy}) should be replaced with $V(\mathbf{x}_e(t),t)$, which is the function of both the position $\mathbf{x}_e(t)$ and the time $t$ due to the Earth rotation, where the subscript $e$ stands for the ECEF frame. In this case, $\partial V(\mathbf{x}_e(t),t)/\partial t$ will not be equal to zero.

The orbit-energy-based recovery of the gravity field of the Earth can be based on either Newtonian mechanics or Hamiltonian mechanics (see e.g., Bjerhammar 1967, 1969; Wolff 1969; Hotine and Morrison 1969; Jekeli 1999; Visser et al. 2003). In what follows, we assume satellite tracking data of positions and velocities in the ECEF and present the principles of orbit-energy-based recovery. First of all, we know the basic relationships of positions and velocities between the ECI and ECEF, which can be readily found in textbooks and are given as follows: \alpheqn \beq \label{ECEF2ECIPosition} \mathbf{x}_i(t) = \mathbf{R}(t)\mathbf{x}_e(t), \eeq \beq \label{ECEF2ECISpeed} \dot{\mathbf{x}}_i(t) = \mathbf{R}(t) ( \dot{\mathbf{x}}_e(t) + \mathbf{\Omega}\times \mathbf{x}_e(t)), \eeq\reseteqn\setcounter{EQ8}{\value{equation}}where $\mathbf{R}(t)$ is the rotation matrix,  $\mathbf{\Omega}$ is the vector of earth rotation  and $\times$ stands for the cross product of two vectors.

Inserting both (\ref{ECEF2ECIPosition}) and (\ref{ECEF2ECISpeed}) into (\ref{satEnergy}), and after some technical arrangement, we finally obtain the constant $E$ in ECI under the framework of the Newtonian mechanics: \beq \label{satEnergyNewton} E = \|\dot{\mathbf{x}}_e(t)\|^2/2 + \Omega (x_{e1}\dot{x}_{e2}-\dot{x}_{e1}x_{e2}) +\Omega^2 (x_{e1}^2+x_{e2}^2)/2 - V(\mathbf{R}(t)\mathbf{x}_e(t)),  \eeq  though $E$ of (\ref{satEnergyNewton}) is expressed in terms of $\dot{\mathbf{x}}_e(t)$ and $\mathbf{x}_e(t)$ in ECEF. We should note that $E$ is conservative in the ECI but not in the ECEF, with the geopotential fixed to the ECI. Alternatively, one may fix the geopotential to ECEF and follow Hamiltonian mechanics (Deshmukh 2019) to recover the gravity field, as is more often used in geodesy (see e.g., Jekeli 1999; Visser et al. 2003). In this case, the Hamiltonian $H$ in ECEF can be written as follows: \alpheqn \beq \label{HamiltonH} H = T - V_{tot}, \eeq where $H$ is the Hamiltonian, which stands for the total energy of the system and is constant, $T$ is the kinetic energy, and $V_{tot}$ is the sum of the gravity and centrifugal potential, namely,
 \beq \label{HamiltonKineticE} T = \|\dot{\mathbf{x}}_e(t)\|^2/2, \eeq, \beq \label{HamiltonPotentialV} V_{tot} = V(\mathbf{x}_e(t)) + \Omega^2 (x_{e1}^2+x_{e2}^2)/2. \eeq\reseteqn\setcounter{EQ8}{\value{equation}}Substituting (\ref{HamiltonKineticE}) and (\ref{HamiltonPotentialV}) into (\ref{HamiltonH}), we obtain the final representation of $H$ as follows:
\beq \label{HamiltonHFinal} H = \|\dot{\mathbf{x}}_e(t)\|^2/2 - V(\mathbf{x}_e(t)) - \Omega^2 (x_{e1}^2+x_{e2}^2)/2. \eeq The Hamiltonian $H$ of (\ref{HamiltonHFinal}) can also be derived through the Lagrangian and Legendre transform, as can be found in Visser et al. (2003). In reality, the Sun, the Moon and planets contribute to the total conservative energy. In this case, these extra contributions should be included in $E$ of (\ref{satEnergy}) or $H$ of (\ref{HamiltonHFinal}) (see e.g., Bjerhammar 1967, 1969; Jekeli 1999). In addition, a satellite in space will also be affected by non-conservative forces such as air-drag, thermal radiation and solar radiation pressure, which should be corrected as well (see e.g., Bjerhammar 1967, 1969; Morrison 1970; Jekeli 1999; Visser et al. 2003).

If we collect a sufficient number of satellite tracking data, we can then estimate the unknown harmonic coefficients $C_{nm}$ and $S_{nm}$ by applying a proper estimation method to the observational equations of type (\ref{satEnergyNewton}) under the Newtonian mechanics or the observational equations of type (\ref{HamiltonHFinal}) under  Hamiltonian mechanics. Since the variables in $V(\mathbf{x}_e(t))$ of the Hamiltonian type (\ref{HamiltonHFinal}) is not involved with the rotation matrix $\mathbf{R}(t)$, it may be more convenient to use (\ref{HamiltonHFinal}) for the recovery of the gravity field.

Because the most important error sources in the orbit-energy-based methods are from the measurement errors of velocities, to fully use the high precision of the range rate between a pair of satellites (Wolff 1969), one can then rewrite the position of the slave satellite in ECEF as: $$ \mathbf{x}_{se}(t) = \mathbf{x}_{me}(t) + \Delta \mathbf{x}(t), $$ from which we have
$$ \dot{\mathbf{x}}_{se}(t) = \dot{\mathbf{x}}_{me}(t) + \Delta \dot{\mathbf{x}}(t), $$
where $\mathbf{x}_{me}(t)$, $\mathbf{x}_{se}(t)$ and $\Delta \mathbf{x}(t)$ are the positions of the master and slave satellites and the relative position between the two satellites, $\dot{\mathbf{x}}_{me}(t)$, $\dot{\mathbf{x}}_{se}(t)$ and $\Delta \dot{\mathbf{x}}(t)$ are the velocity vectors of the master and slave satellites and the relative velocity vector between the two satellites. Subtracting the Hamiltonian $H_s$ of the slave satellite from $H_m$ of the master satellite will eliminate both terms $\|\dot{\mathbf{x}}_{me}(t)\|^2$ and $\|\dot{\mathbf{x}}_{se}(t)\|^2$ but keep only the terms $\dot{\mathbf{x}}^T_{me}(t)\Delta \dot{\mathbf{x}}(t)$ and $\Delta \dot{\mathbf{x}}^T(t)\Delta \dot{\mathbf{x}}(t)$. Therefore, using a pair of satellites for the recovery of the gravity field is advantageous to utilize precise range rates between a pair of satellites. Further technical details and analysis can be found in Jekeli (1999) and will not be presented here. The same procedure can be equally applied to $E$ of (\ref{satEnergy}).

Before closing this subsection, we may like to note that because the orbit-energy-based methods are involved with the mixture of velocity and position measurements, they are incapable of taking the full advantage of the highest  precision of measurements to recover the gravity field, as can be clearly observed either from (\ref{satEnergyNewton}) and (\ref{HamiltonHFinal}) or the corresponding cases of satellite pair. On the other hand, the methods require the most strict condition that both position and velocity of an LEO satellite must be measured simultaneously at each time epoch. Partial information on the orbits and velocities of satellites cannot be directly used in the orbit-energy-based methods. High precision ranges between satellites cannot be well used either. As a result, these methods may not be able to produce high-precision high-resolution gravitational models. This may partly explain why the orbit-energy-based methods receive less attention and are not used to produce standard gravity models either.

\section{Measurement-based perturbation theory and parameter estimation in differential equation }\label{measPert}
Huge technological advances have been made in earth space observation and satellite tracking, especially, since GNSS and satellite gravity missions have been routinely operational. Orbits of LEO satellites can now be tracked almost continuously and have been reported to reach the precision level of 1 cm for each direction (see e.g., \v{S}vehla and Rothacher 2005; Mao et al. 2023). According to the seismo-GNSS experiments by Xu et al. (2013), short term performance of precise point positioning (PPP) can reach the unprecedented precision level of millimeters for all the local East, North and vertical components. This high level of precision may be even more realistic for LEO satellites, since GNSS observation environment in space can be less complex than on the earth surface. In the case of inter-satellite tracking, ranges and range rates are of the precision levels of a few $\mu m$ and $0.1 \mu m/s$ with microwave systems (see e.g., Kim 2000) and even a few $nm$ and  $0.1 nm/s$ with laser ranging interferometers, respectively. The latter is about three orders of magnitude better in accuracy than the former (see e.g., Pierce et al. 2008; Turyshev et al. 2014; Abich et al. 2019).

Satellite gravimetry methods with satellite tracking, as briefly reviewed in the previous sections,  are all limited to short arcs of orbits, physically implying that they are not capable of extracting small gravitational forces from tracking data and further implying that high-precision high-resolution gravitational models are not feasible either. It is very likely that the mathematical modelings with these methods cannot march the extremely high precision of inter-satellite tracking either, as may be inferred from the fitting results of inter-satellite relative orbits in Mao et al. (2023).

The major purpose of this section is to briefly review and summarize the measurement-based perturbation theory for satellite gravimetry proposed by Xu (2008, 2018), which is mathematically convergent uniformly, no matter how lengthy a satellite orbit arc can be. The mathematical uniform convergence is equivalent to saying physically that as far as an orbital arc is continuously tracked precisely and sufficiently lengthy, one will be able to extract any small gravitational forces up to the accuracy limit of satellite tracking observables. Furthermore, because the solution to the Newton's nonlinear differential equations of satellite motion is theoretically free of modeling errors, it is possible to  utilize the full advantage of high precision inter-satellite tracking for gravity recovery as well. Therefore, a high-precision high-resolution gravitational model is feasible and possible, both mathematically and physically. Precise mathematical modeling of satellite tracking observables with the measurement-based perturbation theory may further help refine and gain a deeper insight into non-conservative force models as well.

To start with, let us assume that the orbit of an LEO gravity satellite has been precisely measured or tracked (almost) continuously with GNSS, which is denoted by $\{ \mathbf{x}_m(\tau)| \, 0\leq \tau\leq t\}$ or simply $\mathbf{x}_m(\tau)$. To make our review as concise as possible, with a focus on a precise mathematical solution to the nonlinear differential equations (\ref{NewtonGrav}) (of satellite motion), for simplicity and without loss of generality, we can formally rewrite (\ref{NewtonGrav}) mathematically as follows: \beq \label{NewtonDiffE} \ddot{\mathbf{x}}(t) = \mathbf{f}_1(\mathbf{x}(t),t) + \sum \limits_{n=2}\limits^{N_{max}}\sum \limits_{m=0}\limits^n \{ \mathbf{f}_{nmc}(\mathbf{x}(t), t) C_{nm} + \mathbf{f}_{nms}(\mathbf{x}(t), t) S_{nm}\}, \eeq where the functions $\mathbf{f}_1(\mathbf{x}(t),t)$, $\mathbf{f}_{nmc}(\mathbf{x}(t), t)$ and $\mathbf{f}_{nms}(\mathbf{x}(t), t)$ can all be readily derived from the acceleration functions $\mathbf{a}_E(t, \mathbf{x}(t), \mathbf{p})$ and the disturbing potential $T(r,\mathbf{p})$ through coordinate transformation. The technical derivations of these functions will be omitted here. Actually, one can even directly compute the time functions of longitude, latitude, and radius from $\mathbf{x}_m(t)$. To make our theory to solve nonlinear differential equations with unknown parameters complete and valid for scatteredly tracked data, as was often the case in the early time of satellite gravimetry, we will also briefly summarize the local solutions based on a reference gravity model. For some more technical details on the measurement-based perturbation theory and local solutions, the reader is referred to Xu (2008, 2018).

\subsection{Measurement-based perturbation of positions}
Given approximate initial position $\mathbf{x}_0$ and velocity $\mathbf{v}_0$ with unknown corrections $\delta\mathbf{x}_0$ and $\delta\mathbf{v}_0$, respectively, then we can readily obtain the solution $\mathbf{x}(t)$ to the nonlinear differential equations (\ref{NewtonDiffE}) by applying the double integration to (\ref{NewtonDiffE}) as follows:
 \begin{eqnarray} \label{positionIntDoubleP} \mathbf{x}(t) & = & \mathbf{x}(t_0) + \mathbf{v}(t_0)(t-t_0) + \int_{t_0}^t \int_{t_0}^\eta \mathbf{f}_1(\mathbf{x}(\tau),\tau)d\tau d\eta  \nonumber \\
  &   & + \sum \limits_{n=2}\limits^{N_{max}}\sum \limits_{m=0}\limits^n  C_{nm} \int_{t_0}^t \int_{t_0}^\eta \mathbf{f}_{nmc}(\mathbf{x}(\tau), \tau)d\tau d\eta \nonumber \\
  &   & + \sum \limits_{n=2}\limits^{N_{max}}\sum \limits_{m=0}\limits^n S_{nm}\int_{t_0}^t \int_{t_0}^\eta \mathbf{f}_{nms}(\mathbf{x}(\tau), \tau) d\tau d\eta, \end{eqnarray} where $\mathbf{x}(t_0)=\mathbf{x}_0+\delta\mathbf{x}_0$ and $\mathbf{v}(t_0)=\mathbf{v}_0+\delta\mathbf{v}_0$. For each double integration on the right hand side of  (\ref{positionIntDoubleP}), the integration area or domain is exactly defined by the area $D =\{ (\eta, \tau)\in R^2:\,  t_0\leq\eta\leq t, t_0\leq\tau \leq \eta\}$. If we assume that all the functions are integrable, then according to Fubini theorem, we can change the order of integration over the same area by letting $\tau$ and $\eta$ run over $[t_0,\,t]$ and $[\tau,\,t]$, respectively. As a result, the formal solution (\ref{positionIntDoubleP}) can be equivalently rewritten as follows:
\begin{eqnarray} \label{positionP} \mathbf{x}(t) & = & \mathbf{x}(t_0) + \mathbf{v}(t_0)(t-t_0) + \int_{t_0}^t (t-\tau) \mathbf{f}_1(\mathbf{x}(\tau),\tau)d\tau  \nonumber \\
  &   & + \sum \limits_{n=2}\limits^{N_{max}}\sum \limits_{m=0}\limits^n  C_{nm} \int_{t_0}^t (t-\tau) \mathbf{f}_{nmc}(\mathbf{x}(\tau), \tau)d\tau \nonumber \\
  &   & + \sum \limits_{n=2}\limits^{N_{max}}\sum \limits_{m=0}\limits^n S_{nm}\int_{t_0}^t (t-\tau) \mathbf{f}_{nms}(\mathbf{x}(\tau), \tau) d\tau, \end{eqnarray} (see e.g., Lonseth 1977; Stakgold 2000; Lubansky et al. 2006; Jekeli and Habana 2018; Xu et al. 2021).

Since we have the precise orbital measurements $\mathbf{x}_m(t)$ over the whole orbital arc, we can directly replace $\mathbf{x}(\tau)$  with $\mathbf{x}_m(\tau)$ on the right hand side of (\ref{positionP}) and obtain the first order solution:
\begin{eqnarray} \label{positionP1st} \mathbf{x}_1(t) & = & \mathbf{x}_0 + \mathbf{v}_0(t-t_0) + \int_{t_0}^t (t-\tau) \mathbf{f}_1(\mathbf{x}_m(\tau),\tau)d\tau + \delta\mathbf{x}_0 + \delta\mathbf{v}_0(t-t_0) \nonumber \\
  &   & + \sum \limits_{n=2}\limits^{N_{max}}\sum \limits_{m=0}\limits^n  C_{nm} \int_{t_0}^t (t-\tau) \mathbf{f}_{nmc}(\mathbf{x}_m(\tau), \tau)d\tau \nonumber \\
  &   & + \sum \limits_{n=2}\limits^{N_{max}}\sum \limits_{m=0}\limits^n S_{nm}\int_{t_0}^t (t-\tau) \mathbf{f}_{nms}(\mathbf{x}_m(\tau), \tau) d\tau, \end{eqnarray} (see also Xu 2008, 2018), where $\mathbf{x}_1(t)$ is the first order approximation of $\mathbf{x}(t)$.

Although the true but unknown functions $\mathbf{x}(\tau)$ are required to compute the related terms on the right hand side of (\ref{positionP}), they are only different from $\mathbf{x}_m(\tau)$ by the terms of random measurement errors at any time epoch, namely, $\mathbf{x}(\tau)=\mathbf{x}_m(\tau)+\bm{\epsilon}_x(\tau)$ or equivalently \beq \label{obsPositionXm} \mathbf{x}_m(\tau)=\mathbf{x}(\tau)-\bm{\epsilon}_x(\tau), \eeq where $\bm{\epsilon}_x(\tau)$ is the random measurement noise with zero expectation $E\{\bm{\epsilon}_x(\tau)\} =\mathbf{0}$.  From this point of view, the first order position functions $\mathbf{x}_1(t)$ computed with $\mathbf{x}_m(\tau)$ are different from their true (unknown) values up to the noise level. In fact, by substituting (\ref{obsPositionXm}) into (\ref{positionP1st}) and linearizing it at $\mathbf{x}(\tau)$, we can readily prove
\begin{eqnarray} \label{ExpectpositionP1st} E\{\mathbf{x}_1(t)\} & = & \mathbf{x}(t_0) + \mathbf{v}(t_0)(t-t_0) + \int_{t_0}^t (t-\tau) \mathbf{f}_1(\mathbf{x}(\tau),\tau)d\tau  \nonumber \\
  &   & + \sum \limits_{n=2}\limits^{N_{max}}\sum \limits_{m=0}\limits^n  C_{nm} \int_{t_0}^t (t-\tau) \mathbf{f}_{nmc}(\mathbf{x}(\tau), \tau)d\tau \nonumber \\
  &   & + \sum \limits_{n=2}\limits^{N_{max}}\sum \limits_{m=0}\limits^n S_{nm}\int_{t_0}^t (t-\tau) \mathbf{f}_{nms}(\mathbf{x}(\tau), \tau) d\tau \nonumber \\
  & = & \mathbf{x}(t) \end{eqnarray} up to the first order approximation, because $E\{\bm{\epsilon}_x(\tau)\} =\mathbf{0}$, where $E\{\mathbf{x}_1(t)\}$ stands for the expectation of $\mathbf{x}_1(t)$.
  Therefore, the position solution $\mathbf{x}_1(t)$ of (\ref{positionP1st}) is linearly unbiased, theoretically free of modeling errors up to the first order approximation and capable of utilizing high precision satellite tracking data. It provides a theoretical guarantee for high-precision high-resolution gravitational models
   from satellite and inter-satellite tracking data.

If we require a higher order solution or more precisely, the second order solution, we can apply the Picard method of successive approximation by substituting $\mathbf{x}(t)$ of (\ref{positionP}) with $\mathbf{x}_1(t)$ of (\ref{positionP1st}) and obtain
\begin{eqnarray} \label{positionP2nd} \mathbf{x}_2(t) & = & \mathbf{x}_0 + \mathbf{v}_0(t-t_0) + \int_{t_0}^t (t-\tau) \mathbf{f}_1(\mathbf{x}_1(\tau),\tau)d\tau + \delta\mathbf{x}_0 + \delta\mathbf{v}_0(t-t_0) \nonumber \\
  &   & + \sum \limits_{n=2}\limits^{N_{max}}\sum \limits_{m=0}\limits^n  C_{nm} \int_{t_0}^t (t-\tau) \mathbf{f}_{nmc}(\mathbf{x}_1(\tau), \tau)d\tau \nonumber \\
  &   & + \sum \limits_{n=2}\limits^{N_{max}}\sum \limits_{m=0}\limits^n S_{nm}\int_{t_0}^t (t-\tau) \mathbf{f}_{nms}(\mathbf{x}_1(\tau), \tau) d\tau. \end{eqnarray} An interesting question now is how or where to linearize $\mathbf{f}_1(\mathbf{x}_1(\tau),\tau)$, $\mathbf{f}_{nmc}(\mathbf{x}_1(\tau), \tau)$ and $\mathbf{f}_{nms}(\mathbf{x}_1(\tau), \tau)$ in (\ref{positionP2nd}). If we follow the conventional perturbation theory of small parameters in mathematics, we would immediately treat the small quantity of $\mathbf{x}_1(t)$ in (\ref{positionP1st}), denoted by $\delta\mathbf{x}_1(t)$ and given below
\begin{eqnarray*} \delta\mathbf{x}_1(t) & = & \delta\mathbf{x}_0 + \delta\mathbf{v}_0(t-t_0) \nonumber \\
  &   & + \sum \limits_{n=2}\limits^{N_{max}}\sum \limits_{m=0}\limits^n  C_{nm} \int_{t_0}^t (t-\tau) \mathbf{f}_{nmc}(\mathbf{x}_m(\tau), \tau)d\tau \nonumber \\
  &   & + \sum \limits_{n=2}\limits^{N_{max}}\sum \limits_{m=0}\limits^n S_{nm}\int_{t_0}^t (t-\tau) \mathbf{f}_{nms}(\mathbf{x}_m(\tau), \tau) d\tau \end{eqnarray*} as a small parameter
and then linearize all the terms on the right hand side of (\ref{positionP2nd}) at the best possible approximate solution $\mathbf{x}_1^0(t)$ or more precisely, $$ \mathbf{x}_1^0(t) = \mathbf{x}_0 + \mathbf{v}_0(t-t_0) + \int_{t_0}^t (t-\tau) \mathbf{f}_1(\mathbf{x}_m(\tau),\tau)d\tau. $$ Although a direct application of mathematical perturbation is intuitively attractive, it is not a good practice in our case. The reason is that $\mathbf{x}_1^0(t)$ will get divergent with the increase of time, making the second order solution (\ref{positionP2nd}) useless because of too large a modeling error.

Because the precise orbital measurements $\mathbf{x}_m(\tau)$ are close to their true values along the whole orbital arc, they can be directly used to precisely linearize $\mathbf{x}_2(t)$ of (\ref{positionP2nd}). More specifically, we can linearize $\mathbf{f}_1(\mathbf{x}_1(\tau),\tau)$, $\mathbf{f}_{nmc}(\mathbf{x}_1(\tau), \tau)$ and $\mathbf{f}_{nms}(\mathbf{x}_1(\tau), \tau)$ in (\ref{positionP2nd}) as follows: \alpheqn
\beq \mathbf{f}_1(\mathbf{x}_1(\tau),\tau) = \mathbf{f}_1(\mathbf{x}_m(\tau),\tau) + \mathbf{F}_1(\mathbf{x}_m(\tau),\tau)\{ \mathbf{x}_1(\tau) - \mathbf{x}_m(\tau) \}, \eeq
\beq \mathbf{f}_{nmc}(\mathbf{x}_1(\tau),\tau) = \mathbf{f}_{nmc}(\mathbf{x}_m(\tau),\tau) + \mathbf{F}_{nmc}(\mathbf{x}_m(\tau),\tau)\{ \mathbf{x}_1(\tau) - \mathbf{x}_m(\tau) \}, \eeq
\beq \mathbf{f}_{nms}(\mathbf{x}_1(\tau),\tau) = \mathbf{f}_{nms}(\mathbf{x}_m(\tau),\tau) + \mathbf{F}_{nms}(\mathbf{x}_m(\tau),\tau)\{ \mathbf{x}_1(\tau) - \mathbf{x}_m(\tau) \},
\eeq\reseteqn\setcounter{EQ8A}{\value{equation}}where $$ \mathbf{F}_1(\mathbf{x}_m(\tau),\tau) = \left. \frac{\partial \mathbf{f}_1(\mathbf{x}(\tau),\tau)}{\partial \mathbf{x}} \right|_{ \mathbf{x}(\tau)=\mathbf{x}_m(\tau) }, $$ $$ \mathbf{F}_{nmc}(\mathbf{x}_m(\tau),\tau) = \left. \frac{\partial \mathbf{f}_{nmc}(\mathbf{x}(\tau),\tau)}{\partial \mathbf{x}} \right|_{ \mathbf{x}(\tau)=\mathbf{x}_m(\tau) }, $$
$$ \mathbf{F}_{nms}(\mathbf{x}_m(\tau),\tau) = \left. \frac{\partial \mathbf{f}_{nms}(\mathbf{x}(\tau),\tau)}{\partial \mathbf{x}} \right|_{ \mathbf{x}(\tau)=\mathbf{x}_m(\tau) }. $$
Because all the derivatives are computed at the very precise points and because $[\mathbf{x}_1(\tau) - \mathbf{x}_m(\tau)]$ is a small quantity, there will be no more concern of divergence. Substituting (\arabic{EQ8A}) into (\ref{positionP2nd}) and neglecting all the second order terms except for the term with $C_{20}^2$, we can finally obtain the representation of $\mathbf{x}_2(t)$ with the nonlinear term of $C_{20}^2$. The technical details are omitted here but left to the reader.

\subsection{Measurement-based perturbation of velocities}
To derive the measurement-based perturbation of velocities, we integrate both sides of (\ref{NewtonDiffE}) and obtain the following nonlinear equations
\begin{eqnarray} \label{velocityP} \dot{\mathbf{x}}(t) & = & \mathbf{v}(t_0) + \int_{t_0}^t \mathbf{f}_1(\mathbf{x}(\tau),\tau)d\tau \nonumber \\
  &   & + \sum \limits_{n=2}\limits^{N_{max}}\sum \limits_{m=0}\limits^n  C_{nm} \int_{t_0}^t\mathbf{f}_{nmc}(\mathbf{x}(\tau), \tau)d\tau \nonumber \\
  &   & + \sum \limits_{n=2}\limits^{N_{max}}\sum \limits_{m=0}\limits^n S_{nm}\int_{t_0}^t\mathbf{f}_{nms}(\mathbf{x}(\tau), \tau) d\tau, \end{eqnarray} for the velocity $\dot{\mathbf{x}}(t)$ at the time $t$, where $\mathbf{v}(t_0)$ is the initial velocity of the satellite and has been defined in the above.

In the similar manner to (\ref{positionP1st}), substituting the measured orbital functions $\mathbf{x}_m(\tau)$ into (\ref{velocityP}) yields the first order approximation of the velocity functions, which is also called the quasi-linear approximation in Xu (2008, 2018) and given as follows:
 \begin{eqnarray} \label{velocityP1st} \dot{\mathbf{x}}_1(t) & = & \mathbf{v}_0 + \int_{t_0}^t \mathbf{f}_1(\mathbf{x}_m(\tau),\tau)d\tau + \delta\mathbf{v}_0 \nonumber \\
  &   & + \sum \limits_{n=2}\limits^{N_{max}}\sum \limits_{m=0}\limits^n  C_{nm} \int_{t_0}^t\mathbf{f}_{nmc}(\mathbf{x}_m(\tau), \tau)d\tau \nonumber \\
  &   & + \sum \limits_{n=2}\limits^{N_{max}}\sum \limits_{m=0}\limits^n S_{nm}\int_{t_0}^t\mathbf{f}_{nms}(\mathbf{x}_m(\tau), \tau) d\tau. \end{eqnarray} Although the true but unknown functions $\mathbf{x}(\tau)$ are required to compute the related terms on the right hand side of (\ref{velocityP}), they are essentially only different from $\mathbf{x}_m(\tau)$ by random measurement noise. From this point of view, the first order functions $\dot{\mathbf{x}}_1(t)$ computed with $\mathbf{x}_m(\tau)$ are only different from their true (unknown) values up to the noise level. In fact, by substituting (\ref{obsPositionXm}) into (\ref{velocityP1st}) and linearizing the right hand side of (\ref{velocityP1st}) at $\mathbf{x}(\tau)$, we can also readily prove
 \begin{eqnarray} \label{ExpectvelocityP1st} E\{\dot{\mathbf{x}}_1(t)\} & = & \mathbf{v}(t_0) + \int_{t_0}^t \mathbf{f}_1(\mathbf{x}(\tau),\tau)d\tau  \nonumber \\
  &   & + \sum \limits_{n=2}\limits^{N_{max}}\sum \limits_{m=0}\limits^n  C_{nm} \int_{t_0}^t\mathbf{f}_{nmc}(\mathbf{x}(\tau), \tau)d\tau \nonumber \\
  &   & + \sum \limits_{n=2}\limits^{N_{max}}\sum \limits_{m=0}\limits^n S_{nm}\int_{t_0}^t\mathbf{f}_{nms}(\mathbf{x}(\tau), \tau) d\tau \nonumber \\
  & = & \dot{\mathbf{x}}(t) \end{eqnarray} up to the first order approximation, as is exactly the case of (\ref{ExpectpositionP1st}). Therefore, the solution $\dot{\mathbf{x}}_1(t)$ of (\ref{velocityP1st}) is a linear unbiased estimate of $\dot{\mathbf{x}}(t)$, theoretically free of modeling errors up to the first order approximation and capable of utilizing high precision satellite tracking data. The solution (\ref{velocityP1st}) also provides a theoretical guarantee for high-precision high-resolution gravitational models. If a second order perturbative solution for the velocity of satellite is required, one can follow the same procedures as in (\ref{positionP2nd}) to construct such a solution, which is omitted here as well.

We may note that if position measurements are not sufficiently dense along the orbit, one can first interpolate the positions, follow the above procedures and derive the corresponding perturbation solutions of position and velocity. Both the precise measurement-based perturbation solutions $\mathbf{x}_1(t)$ of (\ref{positionP1st}) and $\dot{\mathbf{x}}_1(t)$ of (\ref{velocityP1st}) and their second order solutions can be readily extended to any satellite. Therefore, one can readily link these solutions to any types of satellite and inter-satellite tracking data such as precise orbital measurements $\mathbf{x}_m(t)$, ranges and range rates to estimate the unknown parameters in the nonlinear differential equations (\ref{NewtonGrav}) of satellite motion, or more precisely, to produce high-precision high-resolution gravitational models from satellite gravimetry.

\subsection{Parameter estimation in differential equations with scattered satellite tracking data}
Although precise and continuous measurement of orbits has become routinely available for LEO satellites, this was unimaginable in the early time of satellite gravimetry. For a certain celestial body such as the Moon, it may still not yet be possible to reach this high level of measurement continuity and accuracy. Thus, we would like to briefly discuss a more general question: how to estimate the unknown parameters in (nonlinear) differential equations, given only a number of scattered (satellite tracking) data over time? More specifically, we assume a reference gravity model and present local solutions to the Newton’s nonlinear governing differential equations of satellite motion for estimating the unknown parameters from scattered data. For more technical details, the reader is referred to Xu (2018).

To start with, in addition to the approximate values $\mathbf{x}_0$ and $\mathbf{v}_0$ of the initial position $\mathbf{x}(t_0)$ and velocity $\mathbf{v}(t_0)$, let us further assume a reference gravity model with the given  approximate values $\mathbf{p}_0$ of the true but unknown parameters $\mathbf{p}$. One option for obtaining  $\mathbf{p}_0$ is to estimate $\mathbf{p}$ using the energy-based methods, if permitted by the measurements.
As a result, we can solve the nonlinear differential equations (\ref{NewtonGrav}) or equivalently (\ref{2ndto1stDiffE}), and obtain the satellite reference trajectories of orbit and velocity, which are denoted by $\mathbf{z}_0(t)$ and satisfy the following differential equations:
 \beq \label{2ndto1stDiffERef} \dot{\mathbf{z}}_0(t) = \left[ \begin{array}{c} \mathbf{v}_0(t) \\ \mathbf{a}_E(t, \mathbf{x}_0(t), \mathbf{p}_0) \end{array} \right] = \mathbf{f}(t, \mathbf{z}_0(t), \mathbf{p}_0). \eeq Subtracting $\dot{\mathbf{z}}_0(t)$ of (\ref{2ndto1stDiffERef}) from $\dot{\mathbf{z}}(t)$ of (\ref{2ndto1stDiffE}) and linearizing the acceleration functions $\mathbf{a}_E(t, \mathbf{x}(t), \mathbf{p})$ at $\mathbf{x}_0(t)$ and $\mathbf{p}_0$, we obtain the linearized differential equations:
 \alpheqn
\beq \label{2ndto1stLinearizedDiffERef} \Delta\dot{\mathbf{z}}(t)  = \left[ \begin{array}{cc} \mathbf{0} & \mathbf{I} \\
     \mathbf{F}_{ax}(t, \mathbf{x}_0(t))  &   \mathbf{0} \end{array} \right] \Delta\mathbf{z}(t) + \left[
     \begin{array}{c} \mathbf{0} \\
     \mathbf{F}_{ap}(t, \mathbf{x}_0(t))  \end{array} \right] \Delta\mathbf{p}, \eeq where $\Delta\mathbf{p}=\mathbf{p}-\mathbf{p}_0$, and
     \beq \label{deltaZt} \Delta\dot{\mathbf{z}}(t) = \dot{\mathbf{z}}(t) - \dot\mathbf{z}_0(t), \eeq
\beq \label{deltaXt} \Delta\mathbf{x}(t) = \mathbf{x}(t) - \mathbf{x}_0(t), \eeq
 \beq \label{deltaVt} \Delta\mathbf{v}(t) = \mathbf{v}(t) - \mathbf{v}_0(t), \eeq
     \beq \label{matrFax} \mathbf{F}_{ax}(t, \mathbf{x}_0(t)) = \left. \frac{\partial \mathbf{a}_E(t, \mathbf{x}, \mathbf{p})}{\partial \mathbf{x}^T}
\right|_{\mathbf{x}=\mathbf{x}_0(t), \,\mathbf{p}=\mathbf{p}_0}, \eeq \beq
\label{matrFap} \mathbf{F}_{ap}(t, \mathbf{x}_0(t)) = \left. \frac{\partial \mathbf{a}_E(t, \mathbf{x}(t),
\mathbf{p})}{\partial \mathbf{p}^T} \right|_{\mathbf{x}=\mathbf{x}_0(t),
\,\mathbf{p}=\mathbf{p}_0}, \eeq\reseteqn\setcounter{EQ9}{\value{equation}}and
$\mathbf{I}$ is a $(3\times 3)$ identity matrix.

Solving the linearized dynamical system (\ref{2ndto1stLinearizedDiffERef}) of differential equations yields the final local solution to (\ref{2ndto1stDiffE}) as follows: \alpheqn
\beq \label{LocalSolutZ} \mathbf{z}(t) = \mathbf{z}_0(t) + \bm{\Phi}(t,t_0)
\Delta\mathbf{z}_0 + \int_{t_0}^t \bm{\Phi}(t,\tau) \left[      \begin{array}{c} \mathbf{0} \\
\mathbf{F}_{ap}(\tau, \mathbf{x}_0(\tau))  \end{array} \right] d\tau \Delta\mathbf{p}, \eeq  in terms of the corrections $\Delta \mathbf{z}_0$  to the initial approximate position $\mathbf{x}_0$ and velocity $\mathbf{v}_0$, and the corrections $\Delta \mathbf{p}$ to the approximate values $\mathbf{p}_0$, where the transition matrix $\bm{\Phi}(t,t_0)$ is defined by the following
differential equations: \beq \label{matrTranEQ}
\dot{\bm{\Phi}}(t,t_0) = \left[ \begin{array}{cc} \mathbf{0} & \mathbf{I} \\
\mathbf{F}_{ax}(t, \mathbf{x}_0(t))  &   \mathbf{0} \end{array} \right] \bm{\Phi}(t,t_0), \eeq (see e.g., Grewal and Andrews 1993; Xu 2018), subject to the initial conditions \beq \label{matrTranConditions} \bm{\Phi}(t_0,t_0) = \mathbf{I}_6, \eeq\reseteqn\setcounter{EQ10}{\value{equation}}with $\mathbf{I}_6$ being a $(6\times 6)$
identity matrix. For more technical details of the local solution (\ref{LocalSolutZ}), the reader is referred to Xu (2018).

Alternatively, assuming the unknown initial conditions $\mathbf{x}(t_0)$ and $\mathbf{v}(t_0)$ with approximate values $[\mathbf{x}_0, \,\mathbf{v}_0]$ and the unknown parameters $\mathbf{p}$ with approximate values $\mathbf{p}_0$, we can also follow the method of Xu (2018) to represent the second local solution at any time epoch in terms of the unknown initial corrections $\Delta\mathbf{z}_0$ to $[\mathbf{x}_0, \,\mathbf{v}_0]$ and the unknown corrections $\Delta\mathbf{p}$ to $\mathbf{p}_0$ by directly solving the nonlinear differential equations (\ref{2ndto1stDiffE}) numerically. For a satellite tracking data $y_i$ at the time epoch $t_{yi}$, to obtain a local solution $\mathbf{z}(t_{yi})$, one may have two options: (i) for all the scattered data collected at the time epochs $\{t_{y0}(=t_0),t_{y1},\,...\,t_{yn}\}$, one may select a most appropriate time interval to numerically represent the local solution $\mathbf{z}(t)$ at a number of time epochs and then interpolate the local solutions at these particular epochs to obtain the representations $\{\mathbf{z}(t_{yi})|\, 1\leq i \leq n\}$ for all the scattered data; (ii) for each $t_{yi}$, one can select an appropriate number of intermediate equal-spaced points to numerically represent $\mathbf{z}(t_{yi})$ in terms of $\Delta\mathbf{z}_0$ and $\Delta\mathbf{p}$. No matter which option is used to derive a local solution, one may use different numerical integration methods such as the Euler method, the extended Euler method, Runge-Kutta methods of any order and/or the Newton-Cotes method; different integration methods can  result in slightly different representations of $\mathbf{z}(t_{yi})$, as can be seen from Xu (2018). Nevertheless, as a general rule for selecting a numerical integration method and, in particular the step size, the accumulated truncation error over the arc should be kept below the observation noise so that the unprecedented accuracy of modern technology of ranging and range-rate measurement can be fully exploited.

We will follow the second option, together with the Euler numerical integration method, to demonstrate how a local solution/representation is constructed in this review. The first step is to divide the time interval $[t_0,\,t_{yi}]$ into $m_{yi}$ equal sub-intervals, namely, $$ t_j = t_0 + j h, \,\, j = 1, 2, ..., m_{yi} $$ where $h=(t_{yi}-t_0)/m_{yi}$. Bearing the reference trajectories $\mathbf{z}_0(t)$ in mind, we can then apply the Euler integration to the nonlinear differential equations (\ref{2ndto1stDiffE}) and obtain the local solution at the next time epoch $t_1$.

To start the Euler method, we have  \beq \label{EulerStart} \mathbf{z}(t_1) =
\mathbf{z}(t_0) +  h\mathbf{f}(t_0, \mathbf{z}(t_0),\mathbf{p}), \eeq (see e.g., Stoer
and Bulirsch 2002; Teodorescu et al. 2013). The next step is to linearize the vector functions
$\mathbf{f}(\cdot)$ at $(\mathbf{z}_0,\, \mathbf{p}_0)$. After some re-arrangement, we can represent $\Delta\mathbf{z}(t_1)$ in terms of $\Delta\mathbf{z}_0$ and $\Delta\mathbf{p}$ as follows:
 \beq \label{EulerStartLin} \Delta\mathbf{z}(t_1) = \delta\mathbf{z}_{01}
+ \left[ \mathbf{I}_6  + h\mathbf{F}_{gz0} \right] \Delta\mathbf{z}_0 +
h\mathbf{F}_{gp0}\Delta\mathbf{p}, \eeq where \alpheqn \beq \delta\mathbf{z}_{01} =
\mathbf{z}_0+ h\mathbf{f}(t_0, \mathbf{z}_0,\mathbf{p}^0) - \mathbf{z}_0(t_1), \eeq
 \beq \label{matrFgz0}
   \mathbf{F}_{gz0} = \left. \frac{\partial \mathbf{f}(t, \mathbf{z}(t),
 \mathbf{p})}{\partial \mathbf{z}^T} \right|_{\mathbf{z}(t)=\mathbf{z}_0,
\,\mathbf{p}=\mathbf{p}_0},  \eeq  and \beq \label{matrFgp0} \mathbf{F}_{gp0} = \left.
\frac{\partial \mathbf{f}(t, \mathbf{z}(t),
 \mathbf{p}) }{\partial \mathbf{p}^T} \right|_{\mathbf{z}(t)=\mathbf{z}_0,
\,\mathbf{p}=\mathbf{p}_0}. \eeq\reseteqn\setcounter{EQ11}{\value{equation}}In a similar manner, we can progress from $t_1$ to $t_2$ by applying the Picard method of successive approximation to (\ref{EulerStart}) and replacing $\mathbf{z}(t_0)$ and $\mathbf{z}(t_1)$ with $\mathbf{z}(t_1)$ and $\mathbf{z}(t_2)$, respectively, and have the local nonlinear representation at $t_2$:
 \beq \label{EulerStartT2} \mathbf{z}(t_2) = \mathbf{z}(t_1) +  h\mathbf{f}(t_1, \mathbf{z}(t_1),\mathbf{p}). \eeq Linearizing (\ref{EulerStartT2}) with some rearrangement yields
\beq \label{EulerStartLinT2} \Delta\mathbf{z}(t_2)  =
\delta\mathbf{z}_{12} + \left[ \mathbf{I}_6  + h\mathbf{F}_{gz1} \right]
\Delta\mathbf{z}(t_1) + h\mathbf{F}_{gp1}\Delta\mathbf{p}, \eeq where
$$ \delta\mathbf{z}_{12} = \mathbf{z}_0(t_1)+ h\mathbf{f}(t_1,
\mathbf{z}_0(t_1),\mathbf{p}^0) - \mathbf{z}_0(t_2). $$  $\mathbf{F}_{gz1}$
and $\mathbf{F}_{gp1}$ are computed after (\ref{matrFgz0}) and (\ref{matrFgp0}) but at the point of $\mathbf{z}_0(t_1)$.

By repeatedly applying the same Picard procedure of successive approximation and linearization as described in the above, we can finally progress to the time epoch $t_{yi}$ of the measurement $y_i$ and obtain its corresponding local solution $\Delta\mathbf{z}(t_{yi})$ as follows: \beq
\label{EulerFinalYi} \Delta\mathbf{z}(t_{yi}) = \delta\mathbf{z}_{0t_{yi}} +
h\sum\limits_{j=1}\limits^{m_{yi}-1}\mathbf{F}_{gzj}\delta\mathbf{z}_{0j} + \left[
\mathbf{I}_6 + h\sum\limits_{j=0}\limits^{m_{yi}-1} \mathbf{F}_{gzj} \right]
\Delta\mathbf{z}_0 + h\sum\limits_{j=0}\limits^{m_{yi}-1}
\mathbf{G}_{gpj}\Delta\mathbf{p}, \eeq as can be found in Xu (2018), where $$ \delta\mathbf{z}_{0t_{yi}} =
\mathbf{z}_0+ h\sum\limits_{j=0}\limits^{m_{yi}-1}\mathbf{f}(t_j,
\mathbf{z}_0(t_j),\mathbf{p}_0) - \mathbf{z}_0(t_{yi}). $$

An alternative way of constructing local solutions to (\ref{NewtonGrav}) is to treat the reference trajectory $\mathbf{z}_0(\tau)\,\,(t_0\leq \tau\leq t)$ as if it were a measured orbit and to derive perturbation solutions of position and velocity after Xu (2008, 2018), which is omitted here. Local solutions such as (\ref{LocalSolutZ}) and (\ref{EulerFinalYi}) can then be used to mathematically describe any scattered satellite tracking data. If there are more than one LEO satellite, one can similarly obtain the solutions of types (\ref{LocalSolutZ}) and/or (\ref{EulerFinalYi}) and use them to construct observational equations for inter-satellite tracking data such as ranges and range rates. Linking the solutions of the Newton's nonlinear differential equations of satellite motion to satellite tracking data is straightforward and will not be further discussed here. We may also like to note that the local solution (\ref{LocalSolutZ}) looks more concise and elegant than (\ref{EulerFinalYi}). However, (\ref{LocalSolutZ}) requires numerically solving for the $(6\times 6)$ transition matrix $\bm{\Phi}(t,t_0)$, while (\ref{EulerFinalYi}) is more direct and may need less computational complexity.

\section{Concluding remarks}
Space geodesy has revolutionized the way we measure the Earth, physically, geometrically and in terms of time variations. It has now become a basic infrastructure for human well-being in modern society and for many different areas of science and engineering. Precise space positioning such as GNSS and InSAR has been most advanced and widely applied, for example, in location, navigation, timing, weather forecasting, space weather, monitoring of tsunami, static and/or kinematic deformation measurement for a variety of structures, and static and/or kinematic deformation monitoring due to natural events such as earthquakes and volcanoes.

Although satellite gravimetry does not influence people’s daily lives as profoundly as GNSS, it has nevertheless made significant progress since the launch of the first artificial satellite in 1957. In particular, thanks to the three dedicated satellite gravity missions, CHAMP, GRACE/GRACE-FO and GOCE, satellite gravimetry has become indispensable for almost all earth-related disciplines, for example, solid earth, hydrology, ocean, sea-level rise, glaciology, climate change and dynamic atmosphere (see e.g., ESA 1996, 1999; NRC 1997;  Dickey 2000; Tapley et al. 2019; Flechtner et al. 2021). However, it faces three critical and fundamental challenges: (i) all standard satellite gravitational models from satellite tracking are routinely computed by using the numerical integration method since 1970s (see e.g., Lerch et al. 1974; Long et al. 1989), which has been proved to be incorrect (Xu 2009, 2018). Although these satellite gravitational products with measurements from CHAMP, GRACE and GRACE-FO missions have been widely applied in many areas of earth science, it is not yet clear how the incorrectness of the method would affect the use and interpretation of these products in all the related areas. Most users of GRACE and GRACE-FO gravity products may not be aware of this fact and probably not able to judge the extent of impact due to this incorrectness of mathematics either. Exactly for these reasons, we make no attempt to review applications of these gravitational products, regardless of how significant their impact may appear; (ii) the resolution of satellite gravity models from GRACE and GRACE-FO still remains low at a scale of 300 to 500 km; and (iii) dedicated satellite gravity missions (CHAMP, GRACE, GRACE-FO and GOCE) are all equipped with accelerometers to measure non-conservative forces. Recent paradigm shift of accelerometers has shown that raw data of acceleration from conventional accelerometers can be numerically incorrect and physically meaningless; actually, accelerometers have been re-defined as a combined system of physical sensing devices and reconstruction algorithms (Xu 2024).

We have provided a unified theoretical framework of parameter estimation in differential equations for satellite gravimetry and briefly reviewed all the major methods to compute gravitational models from satellite and/or inter-satellite tracking, or more precisely, the numerical integration method, the collocation or numerical differentiation-based methods, Kaula linear perturbations, two-point boundary value problems and the orbit-energy-based methods. These classes of methods are either mathematically incorrect, introduce large modeling errors, get divergent and create superficial non-physical phenomena such as critical inclinations and resonance, are limited to short arcs or are not able to take full advantages of continuity of orbits of arbitrary length and unprecedented  accuracy of satellite and/or inter-satellite tracking. As a result, they are physically not capable of extracting small gravitational forces in satellite tracking data, further implying that they are not capable of producing high-precision high-resolution gravitational models. Although the orbit-energy-based methods have often been developed on the basis of Hamiltonian mechanics, we have also expressed the Newton's energy conservation law, with the geopotential fixed to the ECI but calculated with the positions and velocities in the ECEF.

To meet the first two critical and fundamental challenges of satellite gravimetry, we have focused on the most important issue of precise mathematical/functional modelings for satellite and inter-satellite tracking observables and presented the measurement-based perturbation method first proposed by Xu (2008, 2018), which is basically an unbiased solution to the Newton's nonlinear governing differential equations of satellite motion up to random errors and theoretically convergent uniformly over an orbital arc of arbitrary length. Because there is theoretically almost no modeling error, the method is capable of fully utilizing the unprecedented  accuracy of satellite and/or inter-satellite tracking data. Bearing in mind that small gravitational forces will accumulate over time in orbits, the measurement-based perturbation theory is physically capable of extracting such small forces from the continuity of precise orbital arcs of arbitrary length and unprecedented  accuracy of satellite tracking data. From this point of view, high-precision high-resolution gravitational models are feasible and possible, both mathematically and physically, though regularization may add one more aspect of challenge. They should be indeed practically realizable, if LEO satellites of all kinds can be made available to use. As a by-product of high-precision high-resolution gravitational models, filtering, as routinely applied to satellite  gravity fields (see e.g., Wahr et al. 1998; Sasgen et al. 2006; Kusche 2007; Flechtner et al. 2021), may become less important, because it is based on an unrealistic assumption of zero mean and can lead to aliasing effect on or signal distortion of time-varying gravity fields.  Although non-conservative forces can be either measured precisely in the future with computerized accelerometers (Xu 2024) or modelled as part of parameter estimation in differential equations, precise mathematical modeling of satellite tracking observables should help refine and gain a deeper insight into these non-conservative force models. To be complete in theory, we have assumed a reference gravity model and constructed local solutions to the Newton's nonlinear governing differential equations of satellite motion for scattered tracking data that can still be important in some applications, not to mention that this was exactly the case in the early time of satellite gravimetry.

Finally, recognizing that (i) the numerical integration method has been proved to be mathematically incorrect and physically not permitted (Xu 2009, 2018); (ii) almost all standard global gravitational models are produced from satellite and/or inter-satellite tracking on the basis of this method;  and more importantly, (iii) these products have been widely applied in almost all areas of earth science where the users may not be aware of the fact that the products are made on a mathematically incorrect foundation and probably not able to judge the extent of impact due to this incorrectness of mathematics either, we should emphasize that once a mathematical method has been shown to be fundamentally incorrect, continuing to produce and distribute global satellite gravity models based on that method cannot be justified and constitutes a serious scientific issue. This is not merely a matter of implementation but also a matter of scientific responsibility for institutions such as NASA, ESA, GFZ, and CSR (Texas).
\vspace{5mm} \\
{\bf Acknowledgement}: The author thanks Prof. Lars Sj\"{o}berg of KTH very much for his comments on the early history of the energy-based methods for satellite gravimetry. He also thanks one of the reviewers very much for the very constructive comments, which have been fully implemented to substantially improve the manuscript.
\vspace{3mm} \\
{\bf Statement of data availability}: No data is involved in this paper. \\
{\bf  Statement of conflict of interest}: There is no conflict of interest.\\
{\bf  Statement of use of AI}: No AI is used in preparation of this manuscript.\\
{\bf  Statement of fundings}: There is no funding support for this research.
\vspace{5mm} \\
{\bf Notes on contributor:} Peiliang Xu received the Ph.D. degree in geodesy from the
Wuhan Technical University of Surveying and Mapping (now Wuhan University), Wuhan, China, in 1989. He has
been with the Disaster Prevention Research Institute, Kyoto University, Kyoto, Japan, since
1997. He discovers that no accelerometers can produce numerically correct and physically meaningful accelerations in the case of a high sampling rate and provides a patented solution of computerized accelerometers to solve such problems. He has reconstructed the mathematical foundation for tracking-based satellite gravimetry. Recently, he has also found that seismic dynamic displacements are subject to different decay rates in the near- and far-field, which advances precise earthquake early warning on one hand and fails the (universal) empirical scaling law among amplitude, distance and earthquake magnitude with a single decay rate since 1935 on the other hand. His research interests include satellite gravimetry, global navigation satellite system (GNSS)
mathematics and applications, mathematical geodesy, inverse problem theory, statistics/probability and applied
optimization, seismology and civil englineering. Dr. Xu was elected as an IAG Fellow in 2003. He served the International Association of Geodesy (IAG) as its first President of the Intercommission Committee on Theory (2003–2007), the International Union of Geodesy and Geophysics as a member of IUGG Early Career Scientists Award
Committee (2021-2023) and a number of scientific journals as an Associate Editor. He was a recipient of the Best
Paper Award by IAG in 2000 and the Tsuboi Prize by the Geodetic Society of Japan in 2002. He is currently the
Editor-in-Chief of Journal of Geodesy.
\vspace{5mm} \\ {\bf ORCID} \\ 
Peiliang Xu http://orcid.org/0000-0003-1830-8401

\section*{References}
\bl
\ite Abich K, Abramovici A, Amparan B, {\em et al.} (2019). In-orbit performance of the GRACE Follow-on laser ranging interferometer. {\em Phys. Rev. Lett.}, 123, article number 031101.
% has shown biased range measurements
% similar to the primary ranging instrument based on microwaves, but with much less noise at a level of
% 1 nm/Hz^1/2 at Fourier frequencies above 100 mHz (or 0.1 Hz). ==> 10 nm/Hz^1/2 at 40 mHz
\ite Airy GB (1855). On the computation of the effect of the attraction of mountain-masses, as disturbing the apparent astronomical latitude of stations in geodetic surveys. {\em Phil. Trans. R. Soc. London}, 145, 101-104. https://doi.org/10.1098/rstl.1855.0003
 \ite Anderle RJ (1965). Geodetic parameter set NWL-5E-6 based on Doppler satellite
 observations,  NWL Report No.1978, {\em U.S. Naval Weapons Laboratory}, Dahlgren,  Virginia.
 \ite Ballani L (1988). Partielle Ableitungen und Variationsgleichungen zur Modellierung von
Satellitenbahnen und Parameterbestimmung. {\em Vermessungstechnik}, 36, 192-194.
\ite Barnes DF (1966). Gravity changes during the Alaska earthquake. {\em J. geophys. Res.}, 71, 451-456.  https://doi.org/10.1029/JZ071i002p00451
 \ite Benson M (1979). Parameter fitting in dynamic models. {\em Ecol. Modelling}, 6, 97-115.
 \ite Beutler G, J\"{a}ggi A, Mervart L, Meyer U (2010).  The celestial mechanics approach: theoretical foundations. {\em J. Geod.}, 84, 605–624. https://doi.org/10.1007/s00190-010-0401-7
 \ite Bjerhammar A (1967). On the energy integral for satellites. Internal Report of Division of Geodesy, The Royal Institute of Technology, Stockholm.
  \ite Bjerhammar A (1969). On the energy integral for satellites.  {\em Tellus}, 21,   1-9.
 \ite Breiter S, Elipe A (2006). Critical inclination in the main problem of a massive satellite. {\em Celest. Mech. Dyn. Astron.}, 95, 287–297.
 \ite Brewer D, Barenco M, Callard R, Hubank M, Stark J (2008). Fitting ordinary differential equations to short time course data. {\em Phil. Trans. R. Soc. A}, 366, 519-544. https://doi.org/10.1098/rsta.2007.2108
  \ite Brouwer D (1959). Solution of the problem of artificial satellite theory without
 drag. {\em Astron. J.}, 64, 378-396.
  %% artificial satellite motion using standard small parameter perturbation method
  \ite Brouwer D,  Clemence GM (1961). Methods of celestial mechanics.  Academic Press, New York.
  \ite Brumberg VA (1978). Perturbation theory in rectangular coordinates. {\em Celest. Mech.}, 18, 319–336.
\ite Buchar E (1958). Motion of the nodal line of the second Russian Earth satellite (1957$\beta$) and flattening of the Earth. {\em Nature}, 182, 198-199.
\ite Caputo M (1967). The gravity field of the Earth. Academic Press, New York
\ite Chapin D (1998). Gravity instruments: Past, present, future. {\em Leading Edge}, 17, 100-112.
\ite Chen Y, Gu H, Lu Z (1979). Variations of gravity before and after the Haicheng earthquake, 1975, and the Tangshan earthquake, 1976. {\em Phys. Earth Planet. Inter.}, 18, 330-338. \\ https://doi.org/10.1016/0031-9201(79)90070-0
 \ite Coddington EA, Levinson N (1955).  Theory of ordinary differential equations.
McGraw-Hill, New York.
 \ite Coffey SL, Deprit A, Miller BR (1986). The critical inclination in artificial satellite theory. {\em Celest. Mech.},  39, 365–406.
 \ite Comfort G (1974). Direct Mapping of Gravity Anomalies by Using Doppler Tracking between a Satellite Pair. {\em J. geophys. Res.}, 78, 6845-6851.
 \ite Cook AH (1961). Resonant orbits of artificial satellites and longitude terms in the Earth's
  external gravitational potential. {\em Geophys. J. Roy. astr. Soc.}, 4, 53-72.
   \ite Cook AH (1963). The contribution of observations of satellites to the
   determination of the Earth's gravitational potential. {\em Space Sci. Rev.},  2, 355-437.
\ite Cook AH (1965). The absolute determination of the acceleration due to gravity. {\em Metrologia}, 1, 84-114.
\ite Cook AH (1967a). A new absolute determination of the acceleration due to gravity at the National Physical
Laboratory, England. {\em Phil. Trans. R. Soc. London, Ser. A. Math. Phys. Sci.}, 261, 211-252.
 \ite Cook AH (1967b). The determination of the external gravity field of the earth from observations of artificial satellites. {\em Geophys. J. Roy. astr. Soc.}, 13, 297–312. https://doi.org/10.1111/j.1365-246X.1967.tb02161.x
\ite Creutzfeldt B, Güntner A, Thoss H, Merz B, Wziontek H (2010). Measuring the effect of local water storage
changes on in situ gravity observations: Case study of the Geodetic Observatory Wettzell, Germany. {\em Water Resour. Res.}, 46, W08531. https://doi.org/10.1029/2009WR008359
\ite Crossley D, Hinderer J, Riccardi U (2013). The measurement of surface gravity. {\em Rep. Prog. Phys.}, 76, article number 046101. https://doi.org/10.1088/0034-4885/76/4/046101
\ite Dehlinger P (1978). Marine geodesy. Elsevier Scientific Publishing Company, Amsterdam
\ite de Jong K, Scholten R (1973). Gravity and tectonics. John Wiley \& Sons, New York. %% pp.502
\ite Delman A, Landerer F (2022). Downscaling satellite-based estimates of ocean bottom pressure for tracking deep ocean mass transport. {\em Rem. Sens.}, 14(7), 1764. https://doi.org/10.3390/rs14071764
 \ite Deshmukh PC (2019). Foundations of classical mechanics. Cambridge University Press, Cambridge %% ch.6
 \ite Dickey JO (2000). Time variable gravity: an emerging frontier in interdisciplinary geodesy, in: {\em Gravity, Geoid, and Geodynamics 2000}, edited by M Sideris, Springer,  New York, pp.1-5.
  \ite Dickinson RP,  Gelinas RJ (1976). Sensitivity analysis of ordinary
differential equation systems -- a direct method. {\em J. comput. Phys.}, 21, 123-143.
\ite Eckhardt EA (1940). A brief history of the gravity method of prospecting for oil. {\em Geophysics}, 5, 231-242.
\ite E\"{o}tv\"{o}s R (1896). Untersuchungen \"{u}ber Gravitation und Erdmagnetismus. {\em Ann. Physik}, 295, 354-400.
\ite European Space Agency (ESA) (1996). Gravity field and steady-state ocean circulation explorer mission working group. ESA Publication Division, Noordwijk, The Netherlands
\ite European Space Agency (ESA) (1999). Gravity field and steady-state ocean circulation mission. ESA Reports for Mission Selection ESA SP-1233(1), ESA Publication Division, Noordwijk, The Netherlands
\ite Faller JE (1965). Results of an absolute determination of the acceleration of gravity. {\em J. geophys. Res.}, 70, 4035-4038.
\ite Flechtner F, Reigber C, Rummel R, Balmino G (2021). Satellite gravimetry: a review of its realization. {\em Surv. Geophys.}, 42, 1029–1074.
\ite Florsch N, Hinderer J (2000). Bayesian estimation of the free core nutation parameters from the
analysis of precise tidal gravity data. {\em Phys. Earth Planet. Inter.}, 117, 21–35.
  \ite Gaposchkin EM (1974). Earth's gravity field to the eighteenth degree and geocentric coordinates
  for 104 stations from satellite and terrestrial data. {\em J. geophys. Res.}, 79, 5377-5411.
 \ite Gerlach C, Sneeuw N, Visser P, \v{S}vehla D (2003). CHAMP gravity field recovery using the energy balance approach. {\em Adv. Geosci.}, 1, 73–80. https://doi.org/10.5194/adgeo-1-73-2003.
  \ite Gou J, B\"{o}rger L, Schindelegger M, Soja B (2025). Downscaling GRACE-derived ocean bottom pressure anomalies using self-supervised data fusion. {\em J. Geod.}, 99, article number 19. \\ https://doi.org/10.1007/s00190-025-01943-9
   \ite Grewal MS, Andrews AP (1993). Kalman filtering.  Prentice Hall, New Jersey.
\ite Grewal MS, Weill LR, Andrews AP (2001) Global positioning systems, inertial
 navigation and integration, Wiley, New York.
\ite  Gronwall TH (1919). Note on the derivatives with respect to a parameter of the solutions
 of a system of differential equations. {\em Ann. Math.}, 20, 292-296.
 \ite Groves GV (1960). Motion of a satellite in the Earth’s gravitational field. {\em Proc. Roy. Soc. Lond.},  A254, 48–65.
\ite Guier WH, Newton RR (1965). The Earth's gravity field as deduced from the doppler tracking of five satellites. {\em J. geophys. Res.}, 70, 4613-4626.
 \ite Hagihara Y (1972). Celestial mechanics, Vol.II, Part 1: Perturbation theory. MIT
 Press, Cambridge.   %% disturbing potential: extremely small masses page 129
 \ite Hammer S (1945). Estimating ore masses in gravity prospecting. {\em Geophysics}, 10, 50–62.
 \ite Han S, Shum CK, Bevis M, Chen J, Kuo C (2006). Crustal dilatation observed by GRACE after the 2004 Sumatra-Andaman Earthquake. {\em Science}, 313, 658-662. %% energy method
\ite Hansen E (1992). Global optimization using interval analysis. Marcel Dekker, New York.
 \ite Hecker O (1903). Bestimmung der Schwerkraft auf dem atlantischen Ozean. Ver\"{o}ffentlichung des K\"{o}nigl Preuszischen geod\"{a}tischen Institutes, Neue Folge No.11, Verlag von P. Stankiewicz' Buchdruckerei, Berlin
 \ite Heiskanen WA, Moritz H (1967). Physical geodesy.  W.H. Freeman Publ. Co., San Francisco.
  \ite Hotine M, Morrison F (1969). First integrals of the equations of satellite motion.
 {\em Bull. Geod.}, 43, 41-45.
\ite Howarth RJ (2007). Gravity surveying in early geophysics. I. From time-keeping to figure of the Earth. {\em Earth Sci. Hist.}, 26, 201–228.
 \ite  Howland JL, Vaillancourt R (1961). A generalized curve-fitting procedure, {\em SIAM J.
 appl. Math.}, 9, 165-168.  %%% using the sensitivity analysis method
 \ite Hu T, Qiu YP, Cui HJ, Chen LH (2015). Numerical discretization-based kernel type estimation methods for ordinary differential equation models. {\em Acta Math. Sinica English Series}, 31, 1233–1254.
DOI: 10.1007/s10114-015-4256-y
   \ite  Hwang J-T, Dougherty EP, Rabitz S, Rabitz H (1978).  The Green's function
method of sensitivity analysis in chemical kinetics, {\em J. chem. Phys.}, 69, 5180-5191.
  \ite Ilk KH, Feuchtinger M, Mayer-G\"{u}rr T (2005). Gravity field recovery and validation by analysis of short arcs of a satellite-to-satellite  tracking experiment as CHAMP and GRACE. in: Sanso F. ed. {\em A Window on the
 Future of Geodesy}, pp.189-194, Springer, Berlin.
\ite Ilk KH,  L\"{o}cher A, Mayer-G\"{u}rr T (2008). Do we need new gravity field recovery techniques for the new gravity field satellites? In: Xu P.L., Liu J.N. \& Dermanis A. eds. {\em Proc. VI Hotine-Marussi Symp. theor. comput. Geodesy}, pp.3-8, Springer, Berlin.
\ite Izsak IG (1963). Tesseral harmonics in the geopotential. {\em Nature}, 199, 137–139.
 \ite J\"{a}ggi A, Dahle C, Arnold D, Bock H, Meyer U, Beutler G, van den IJssel J (2016). Swarm kinematic orbits and gravity fields from 18 months of GPS data. {\em Adv. Space Res.}, 57, 218-233.
\ite Jekeli C (1993). A review of gravity gradiometer survey system data analyses. {\em Geophysics}, 58, 508-514.
    \ite Jekeli C (1999). The determination of gravitational potential differences from satellite-to-satellite tracking. {\em Celest. Mech. Dynam.  Astron.}, 75, 85-101.
 \ite Jekeli C, Habana N (2018). On the numerical implementation of a perturbation method for satellite gravity mapping. Presented at IX Hotine-Marussi Symposium Rome, 18–22 June 2018
\ite Jiang X, Guo J, Lin M, Sun H, Jiang T (2024). Enhanced gravity-geologic method to predict bathymetry by considering non-linear effects of surrounding seafloor topography. {\em Geophys. J. Int.}, 239, 754–767. https://doi.org/10.1093/gji/ggae301
 \ite Jupp AH (1988). The critical inclination problem – 30 years of progress. {\em Celest. Mech.}, 43, 127–138.
\ite Kaula WM (1963). Determination of the Earth's gravitational field. {\em Rev. Geophys.}, 1, 507-551.
\ite Kaula WM (1966). Theory of satellite geodesy. Blaisdell Publishing Company, London
  \ite Kim J  (2000). Simulation study of a low-low satellite-to-satellite
tracking missions. {\em PhD Dissertation}, The University of Texas at Austin.
% a high accuracy microwave raging system
\ite King-Hele DG, Merson RH (1959). A new value of the Earth's flattening, derived from measurements of satellite orbits.  {\em Nature}, 183, 881-882.
  \ite Kloko\v{c}n\'{i}k J, Gooding RH, Wagner CA, Kosteleck\'{y} J,  Bezd\v{e}k A  (2013). The use of resonant orbits in satellite geodesy: A review. {\em Surv. Geophys.},  34, 43-72.
\ite Kotsakis C, Sideris MG (1999). On the adjustment of combined GPS/levelling/geoid networks. {\em J. Geod.}, 73, 412–421.
 \ite Kozai Y (1959). The motion of a close Earth satellite. {\em Astron. J.}, 64, 367-377.
 \ite Kozai Y (1962). Second-order solution of artificial satellite theory  without air drag, {\em Astron. J.}, 67, 446-461.
 \ite Kusche J (2007). Approximate decorrelation and non-isotropic smoothing of time-variable GRACE-type gravity field models. {\em J. Geod.}, 81, 733–749.
\ite LaCoste LJB (1934). A new type long period vertical seismograph. {\em Physics}, 5, 178-180.
\ite LaCoste LJB (1967). Measurement of gravity at sea and in the air. {\em Rev. Geophys.}, 5, 477-526.
\ite LaCoste L (1988). The zero-length spring gravity meter. {\em Leading Edge}, 7, 20–21.
  \ite Lambeck K, Coleman R (1983). The Earth's shape and gravity field: a report of  progress from 1958 to 1982. {\em Geophys. J. Roy. astr. Soc.}, 74, 25-54.
 \ite Law VJ, Sharma Y (1997). Computation of the gradient and sensitivity coefficients in sum of squares minimization problems with differential equation models. {\em Comput. chem. Eng.}, 21, 1471-1479.
\ite Lenzen VF, Multhauf RP (1966). Development of gravity pendulums in the 19th century. {\em United States National Museum Bulletin}, 240, Paper 44, Contributions from The Museum of History and Technology, Smithsonian Institution. Washington
 \ite Lerch FJ, Wagner CA, Richardson JA, Brownd JE (1974). Goddard Earth Models (5 and
 6).  Technical Report NASA-TM-X-70868, Goddard Space Flight Center, Maryland.
  \ite Liang H, Wu H  (2008). Parameter estimation for differential equation models using a framework
  of measurement error in regression models. {\em J. Amer. statist. Ass.}, 103,
  1570-1583.
   \ite  Linga P, Al-Saifi N, Englezos P (2006). Comparison of the Luus-Jaakola optimization and
 gauss-newton methods for parameter estimation in ordinary differential equation models, {\em Ind. Eng. Chem. Res.}, 45, 4716-4725.  %%% simply follow the incorrect sensitivity method with S(t0, p) = 0
 \ite Long AC, Cappellari JO, Velez CE, Fuchs AJ (1989). Goddard trajectory determination system (GTDS) mathematical theory. Technical Report FDD/552-89/0001 and  CSC/TR-89/6001,
 Goddard Space Flight Center, Maryland.
 \ite Lonseth AT (1977). Sources and applications of integral equations. {\em SIAM Rev.}, 19, 241–278.
 \ite Lubansky AS, Yeow YL, Leong Y-K, Wickramasinghe SR, Han B (2006). A general method of computing the derivative of experimental data. {\em AIChE J.}, 52, 323–332.
 \ite Mao X, Arnold D, Kalarus M, Padovan S, J\"{a}ggi A (2023). GNSS-based precise orbit determination for maneuvering LEO satellites. {\em GPS Solut.}, 27, article number 147.
% Internally, in each direction, agreement between the reduced-dynamic and kinematic orbits reaches a level
% of 1 cm; SLR and K-band ranging measurements indicate a 1-cm accuracy of absolute orbits and
% a 2-mm accuracy of the GRACE-FO relative orbits
\ite Matthews MR (2014). Pendulum motion: A case study in how history and philosophy can contribute to science education. In: {\em International Handbook of Research in History, Philosophy and Science Teaching}, M.R. Matthews (ed.), pp.19-56, Springer, New York
  \ite Mayer-G\"{u}rr T, Ilk KH, Eicker A,  Feuchtinger M (2005). ITG-CHAMP01: a CHAMP gravity field model from short kinematic arcs over a  one-year observation period. {\em J. Geod.}, 78, 462-480.
 \ite Milbert D, Jekeli C (2023). On the initialization of the sensitivity matrix in variational equations. {\em J. Geod.}, 97, article number 88.
\ite Milsom J (2018). The hunt for Earth gravity: A history of gravity measurement from Galileo to the 21st century. Springer International Publishing AG, Cham
 \ite Moles CG, Mendes P, Banga JR (2003). Parameter estimation in biochemical pathways: A comparison of global optimization methods. {\em Genome Res.}, 13, 2467–2474.
 \ite Morrison F (1970). Comments on Paper by Milo Wolff, `Direct Measurements of the Earth's Gravitational Potential Using a Satellite Pair'. {\em J. geophys. Res.}, 75, 2142-2143.
 \ite Musen P, Carpenter L (1963). On the general planetary perturbations in rectangular coordinates. {\em J. geophys. Res.}, 68, 2727-2734.
\ite National Research Council (NRC) (1997). Satellite gravity and the geosphere: Contributions to the study of the solid earth and its fluid envelopes. National Academy Press, Washington, D.C.
\ite Nettleton LL, LaCoste LJ, Harrison JC (1960). Tests of an airborne gravity meter. {\em Geophysics}, 25, 181–202.
\ite O'Keefe JA, Eckeis A, Squires RK (1959). Vanguard measurements give pear-shaped component of Earth's figure. {\em Science}, 129, 565-566.
\ite Okubo S (2020). Advances in gravity analyses for studying volcanoes and earthquakes. {\em Proc. Japan Acad. B}, 96, 50-69. https://doi.org/10.2183/pjab.96.005
 \ite Pavlis NK, Holmes SA, Kenyon SC, Factor JK (2012). The development and evaluation of the Earth
Gravitational Model 2008 (EGM2008). {\em J. geophys. Res.}, 117, B04406. https://doi.org/10.1029/2011JB008916
 \ite Peifer M, Timmer J (2007). Parameter estimation in ordinary differential equations for biochemical processes using the method of multiple shooting. {\em IET Syst. Biol.}, 1, 78-88.
 \ite Peterson TI (1962). Kinetics and mechanism of naphthalene oxidation by non-linear estimation. {\em Chem.  Eng. Sci.}, 17, 203-219.
  \ite Pierce R, Leitch J, Stephens M, Bender P, Nerem R (2008). Intersatellite range monitoring
 using optical interferometry. {\em Appl. Opt.}, 47, 5007-5018.
 % an interferometer designed to provide 1-10nm/Hz^1/2 displacement measurement resolution,
 % in the range 0.01 Hz to 1 Hz
\ite Poynting JH, Thomson JJ (1907). A textbook of physics. Charles Griffin and Company Limited, Strand
\ite Pratt JH (1855). On the attraction of the Himalaya Mountains, and of the elevated regions beyond them, upon the plumb-line in India. {\em Phil. Trans. R. Soc. London}, 145, 53-100. \\ https://doi.org/10.1098/rstl.1855.0002
\ite Ramberg H (1967). Gravity, deformation and the Earth's crust. Academic Press, London
 \ite Reigber C (1989). Gravity field recovery from satellite tracking data. In: Sanso F, Rummel R, editors. {\em Theory of satellite geodesy and gravity field determination}. Lecture notes in Earth Sciences, 25, Berlin,  Springer, pp.197–234.
 \ite Reubelt T, Austen G, Grafarend E (2003). Harmonic analysis of the Earth’s gravitational field by means of
semi-continuous ephemerides of a low Earth orbiting GPS-tracked satellite. Case study: CHAMP.
{\em J. Geod.}, 77, 257–278.
\ite Riley JD, Bennett MM, McCormick E (1967). Numerical integration
of variational equations. {\em Math. Comput.}, 21, 12-17.
 \ite Ritt JF (1919). On the differentiability of the solution of a differential equation
 with respect to a parameter. {\em Ann. Math.}, 20, 289-291.
 %% differential equations for the unknown parameter
  \ite Rummel R (1986). Satellite gradiometry, In: {\em Mathematical and Numerical
Techniques in Physical Geodesy}, edited by H S\"{u}nkel, Springer, Berlin, pp.317-363.
\ite Rummel R, Rapp RH (1977). Undulation and anomaly estimation using Geos-3 altimeter data without precise satellite orbits. {\em Bull. Geod.}, 51, 73–88.
 \ite Rummel R, Yi WY,  Stummer C (2011). GOCE gravitational gradiometry.  {\em J. Geod.}, 85, 777-790.
 \ite Sarode KD, Kumar VR, Kulkarni BD (2015) Embedded multiple shooting methodology in a genetic algorithm framework for parameter estimation and state identification of complex systems. {\em Chem. Eng. Sci.}, 134, 605–618.
 \ite Sasgen I, Martinec Z, Fleming K (2006). Wiener optimal filtering of GRACE data. {\em Stud. Geophys. Geod.},  50, 499–508.
 \ite Schneider M (1968). A general method of orbit determination. Report 1279, Royal Aircraft Establishment, Hants, England
 \ite Schneider M (1984). Observation equations based on expansions into eigenfunctions. {\em Manuscr. Geod.}, 9,
169–208.
 \ite Schneider M (2006). Gravitationsfeldbestimmung unter Verwendung von Bilanzgleichungen f\"{u}r beliebige Observablen. IAPG FESG No.23, Inst. Astron. Phys. Geod., M\"{u}nchen.
\ite Schrama EJO (1989). The role of orbit errors in processing of satellite altimeter data.  Neth. Geod. Comm. Publ., New Ser. No.33, Rijkscomm. Geod., Delft
 \ite Scitovski R, Juki\'{c} D (1996). A method for solving the parameter identification problem for ordinary differential equations of the second order. {\em Appl. Math. Comput.}, 74, 273-291.
   \ite Seeber G (2003). {\em Satellite Geodesy}, 2nd edn., Walter de Gruyter,  Berlin.
 \ite Stakgold I (2000). Boundary value problems of mathematical physics. vol.1, SIAM, Philadelphia.
\ite Stoer J, Bulirsch R (2002). Introduction to numerical analysis. 3rd edn, Springer, Berlin.
\ite Stokes GG (1849). On the variation of gravity at the surface of the earth. {\em Trans. Cambridge Phil. Soc.}, 8, 672-695.
    \ite \v{S}vehla D, Rothacher M (2005). Kinematic positioning of LEO and GPS satellites and IGS stations on the ground. {\em Adv. Space Res.}, 36, 376-381.
\ite Taff LG (1985). Celestial mechanics: A computational guide for the practitioners. Wiley-Interscience, New York.
  \ite Tanaka Y, Okubo S, Machida M, Kimura I, Kosuge T (2001). First detection of absolute gravity change caused by earthquake. {\em Geophys. Res. Lett.}, 28, 2979-2981.
\ite Tapley BD, Watkins MM, Flechtner F, et al. (2019). Contributions of GRACE to understanding climate change. {\em Nat. Clim. Chang.}, 9, 358–369. https://doi.org/10.1038/s41558-019-0456-2
 \ite Teodorescu P, St\u{a}nescu N, Pandrea N (2013).  Numerical analysis with applications in mechanics and engineering, Wiley, New Jersey.
 \ite Tjoa I,  Biegler LT (1991). Simultaneous solution and optimization strategies for parameter estimation of differential-algebraic equation systems. {\em Ind. Eng. Chem. Res.}, 30, 376-385.
\ite Tomoda Y (2010). Gravity at sea -- A memoir of a marine geophysicist. {\em Proc. Jpn. Acad. Ser. B}, 86, 769-787.
\ite Torge W (2001). Geodesy, 3rd Ed. Walter de Gruyter, Berlin
\ite Tsuboi C, Fuchida T (1937). Relations between gravity values and corresponding subterranean mass distribution.  {\em Bull. Earthq. Res. Inst., Tokyo Imperial University}, 15, 636-649.
   \ite Turyshev SG, Sazhin MV, Toth VT (2014). General relativistic laser interferometric
  observables of the GRACE-Follow-On mission. {\em Phys. Rev. D}, 89, article number 105029.
% We develop a relativistic model for the LRI-enabled range between the two GRACE-FO spacecraft, accurate to
% less than 1 nm, and a high-precision model for the corresponding range rate, accurate to better than 0.1 nm/s.
 \ite Van den Bosch B, Hellinckx L (1974). A new method for the estimation of parameters in differential equations. {\em AlChE J.}, 20, 250-255
 \ite Van Domselaar B, Hemker PW (1975). Nonlinear parameter estimation in initial value problems. Report NW 18/75, Math. Center, Amsterdam
\ite Varah  JM (1982). A spline least squares method for numerical parameter estimation in differential equations. {\em SIAM J. sci. stat. Comput.}, 3, 28-46.
\ite Vening Meinesz FA (1929). Theory and practice of pendulum observations at sea.  Publ. Neth. Geod. Comm., Techn Boekhandel Druk. Delft
 \ite Visser P, Sneeuw N, Gerlach C (2003). Energy integral method for gravity field determination
from satellite orbit coordinates. {\em J. Geod.}, 77, 207-216.
 \ite Wahr J, Molenaar M, Bryan F (1998). Time variability of the Earth's gravity field: Hydrological and oceanic effects and their possible detection using GRACE. {\em J. geophys. Res.}, 103(B12), 30205–30229.
 \ite  Wang B, Enright W (2013). Parameter estimation for odes using across-entropy
 approach, {\em SIAM J. sci. Comput.}, 35, A2718-A2737.
 %%% simply follow the incorrect sensitivity method with S(t0, p) = 0
   \ite Wolff M (1969). Direct measurements of the Earth's gravitational  potential using a satellite pair. {\em J. geophys. Res.}, 74, 5295-5300.  %% up to 2000, 37 cittaions, 2001-2026, 126 citations
 \ite Wunsch C, Gaposchkin EM (1980).  On using satellite altimetry to determine the general circulation
of the oceans with application to geoid improvement. {\em Rev. Geophys. Space Phys.}, 18, 725-745.  https://doi.org/10.1029/RG018i004p00725
\ite Xu H, Yu J, Zeng Y, Wang Q, Tian Y, Sun Z (2024). Predicting bathymetry based on vertical gravity gradient anomaly and analyses for various influential factors. {\em Geodesy Geodyn.}, 15, 386-396.
\ite Xu PL (2003). A hybrid global optimization method: The multidimensional case. {\em J. comput. appl.  Math.}, 155, 423-446.
 \ite Xu PL (2008). Position and velocity perturbations for the determination of geopotential from space geodetic measurements. {\em Celest. Mech. dynam. Astron.},   100, 231-249.
\ite Xu PL (2009). Zero initial partial derivatives of satellite orbits with respect
to force parameters violate the physics of motion of celestial bodies, {\em Science China
Series D: Earth Sci.}, 52, 562-566.
 \ite Xu PL (2018). Measurement-based perturbation theory and differential equation parameter
estimation with applications to satellite gravimetry. {\em Commun. Nonlinear
Sci. Numer. Simulat.}, 59, 515-543. https://doi.org/10.1016/j.cnsns.2017.11.021
  \ite Xu PL (2019). Improving the weighted least squares estimation of parameters in
 errors-in-variables models.  {\em J. Frankl. Inst.-Eng. Appl. Math.}, 356, 8785-8802.
 https://doi.org/10.1016/j.jfranklin.2019.06.016
   \ite Xu PL (2024). Concept of computerized accelerometers. {\em Sens. Actuator A-Phys.}, 378, article number 115787. https://doi.org/10.1016/j.sna.2024.115787
  \ite Xu PL, Shi C, Fang RX, Liu JN, Niu XJ, Zhang Q, Yanagidani T (2013).  High-rate precise point positioning (PPP) to measure seismic wave motions: An experimental comparison of GPS PPP with inertial
measurement units. {\em J. Geod.},  87, 361-372. https://doi.org/10.1007/s00190-012-0606-z
  \ite Xu PL, Liu JN, Zeng W, Shen YZ (2014). Effects of errors-in-variables on weighted
 least squares estimation. {\em J. Geod.}, 88, 705-716. https://doi.org/10.1007/s00190-014-0716-x
   \ite Xu PL, Du F, Shu Y, Zhang HP, Shi Y (2021). Regularized reconstruction of peak ground velocity and acceleration from very high-rate GNSS precise point positioning with applications to the 2013 Lushan Mw6.6 earthquake. {\em J. Geod.}, 95, article number 17. https://doi.org/10.1007/s00190-020-01449-6
\ite Yokoyama I (1989). Microgravity and height changes caused by volcanic activity: four Japanese examples. {\em Bull. Volcanol.}, 51, 333-345.
\el

\end{document}